\documentclass[lettersize,journal]{IEEEtran}
\usepackage{algorithm}
\usepackage{algorithmicx}
\usepackage{algpseudocode}
\usepackage{amsmath,amsfonts}
\usepackage{array}
\usepackage{subfig}
\usepackage{textcomp}
\usepackage{enumitem}
\usepackage{stfloats}
\usepackage{url}
\usepackage{verbatim}
\usepackage{graphicx}
\usepackage{bm}
\usepackage{svg}
\usepackage{cite}
\usepackage{subcaption}
\usepackage{threeparttable}
\usepackage{multirow}
\makeatletter
\newcommand{\rmnum}[1]{\romannumeral #1}
\newcommand{\Rmnum}[1]{\expandafter@slowromancap\romannumeral #1@}
\makeatother
\begin{document} 

\title{A Digital Twin-based Intelligent Network Architecture for Underwater Acoustic Sensor Networks}
\author{Shanshan Song, ~\IEEEmembership{Member, IEEE},
  Bingwen Huangfu~\IEEEmembership{},
  Jiani Guo*~\IEEEmembership{},
  Jun Liu, ~\IEEEmembership{Member, IEEE},\\
  Junhong Cui,
  and Xuemin (Sherman) Shen,~\IEEEmembership{Fellow, IEEE}
  \thanks{J. Guo is the corresponding author.}
  \thanks{S. Song, B. Huangfu, and J. Guo are with the College of Computer Science and Technology, Jilin University, Changchun 130012, China (e-mail: songss@jlu.edu.cn, hfbw24@mails.jlu.edu.cn, jnguo20@mails.jlu.edu.cn).}
  \thanks{J. Liu is with the School of Electronic and Information Engineering, Beihang University, Beijing 100191, China (e-mail: liujun2019@buaa.edu.cn).}
  \thanks{J. Cui is with Shenzhen Institute for Advanced Study, UESTC, Shenzhen, China, and also with the College of Computer Science and Technology, Jilin University, Changchun, China (e-mail: junhong\_cui@jlu.edu.cn).}
  \thanks{Xuemin (Sherman) Shen is with the Department of Electrical and Computer Engineering, University of Waterloo, Waterloo, ON, Canada, N2L 3G1 (e-mail: sshen@uwaterloo.ca).}
}

\markboth{Journal of \LaTeX\ Class Files,~Vol.~14, No.~8, August~2021}%
{Shell \MakeLowercase{\textit{et al.}}: A Sample Article Using IEEEtran.cls for IEEE Journals}

\maketitle

\begin{abstract}

  Underwater acoustic sensor networks (UASNs) drive toward strong environmental adaptability, intelligence, and multifunctionality. However, due to unique UASN characteristics, such as long propagation delay, dynamic channel quality, and high attenuation, existing studies present untimeliness, inefficiency, and inflexibility in real practice. Digital twin (DT) technology is promising for UASNs to break the above bottlenecks by providing high-fidelity status prediction and exploring optimal schemes.
  In this article,
  we propose a Digital Twin-based Network Architecture (DTNA),
  enhancing UASNs' environmental adaptability, intelligence, and multifunctionality. By extracting real UASN information from local (node) and global (network) levels, we first design a layered architecture to improve the DT replica fidelity and UASN control flexibility. In local DT, we develop a resource allocation paradigm (RAPD), which rapidly perceives performance variations and iteratively optimizes allocation schemes to improve real-time environmental adaptability of resource allocation algorithms. 
  In global DT, we aggregate decentralized local DT data and propose a collaborative Multi-agent reinforcement learning framework (CMFD) and a task-oriented network slicing (TNSD). 
  CMFD patches scarce real data and provides extensive DT data to accelerate AI model training.
  TNSD unifies heterogeneous tasks' demand extraction and efficiently provides comprehensive network status, improving the flexibility of multi-task scheduling algorithms. 
  Finally, practical and simulation experiments verify the high fidelity of DT. Compared with the original UASN architecture, experiment results demonstrate that DTNA can: 
  (\rmnum{1}) improve the timeliness and robustness of resource allocation;
  (\rmnum{2}) greatly reduce the training time of AI algorithms;
  (\rmnum{3}) more rapidly obtain network status for multi-task scheduling at a low cost.

  \end{abstract}
  
  
  \section{Introduction}
  
  Underwater Acoustic Sensor Networks (UASNs) have been extensively utilized in a variety of scenarios, such as auxiliary navigation, disaster warning, and unmanned exploration \cite{adhoc,Pollution,towards,Unified}.	
  Given the evolving research on underwater applications, 
  there is a desire for UASNs to possess strong environmental adaptability, intelligence, and multifunctionality.
  To achieve the above objectives, most of existing studies are focused on resource allocation, AI algorithms, and multi-task scheduling.
  However, constrained by the unique underwater acoustic channel characteristics, UASNs barely satisfy these studies' requirements for timeliness, efficiency, and flexibility \cite{uchannel} \cite{IOUT}. We summarize various issues of the current UASNs in the development of resource allocation, AI algorithms, and multi-task scheduling as follows: 
  
  \textit{Allocating appropriate resources lacks robustness and timeliness.}
  Underwater acoustic channel exhibits temporal and spatial variability \cite{eli}. 
  To adapt to such a dynamic underwater environment, allocating appropriate resources is essential to improve node and network performance.
  However, UASNs' unreliable communication limits nodes to obtain comprehensive channel and neighboring information, leading to deviated resource allocation schemes. Directly implementing such schemes reduce the UASNs' robustness. 
  To rectify the deviated scheme, traditional UASNs require multiple adjustments to iterate the optimal solution.
  However, due to long propagation delay, the adjustments require long-term communication, leading to resource allocation untimeliness.
  
  \textit{Multi-agent reinforcement learning (MARL) training lacks efficiency.}
  MARL is a promising approach for addressing a variety of underwater tasks, such as multi-autonomous underwater vehicle (AUV) path planning, collaborative power allocation, and sensor data collection \cite{mappo,Self,ACH}.
  However, due to long propagation delay and low transmission rate, UASNs wait for a long time to receive the corresponding reward after each action, resulting in low collection efficiency of training data.
  UASN nodes typically obtain observations through communication, but the high bit error rate (BER) of acoustic communication can impair observations and thereby severely affect the action selection.
  Moreover, the training data with impaired observations are invalid for MARL algorithms, further exacerbating the data scarcity. 
  In addition, long training time indicates extensive data communications that consume excessive energy, reducing the network lifetime.
  
  \textit{Multi-task scheduling lacks flexibility.}
  As the demand for underwater applications grows, one network needs to serve multiple tasks.
  Nowadays, researchers typically deploy large-scale, node-heterogeneous networks to tackle multi-task scenarios \cite{Task}.
  To simplify multi-task scheduling and reduce inter-task conflicts, existing methods normally divide one network into various subnets, and each subnet performs an individual task.
  However, such an approach rigidifies the network topology, and each node is constrained into a preconfigured subnet.
  Under such a condition, a subnet may have idle nodes with excess energy while nodes in another subnet may exhaust all energy, which reduces the task efficiency and leads to unfair multi-task scheduling.
  An alternative solution is to design a centralized scheduling, but it requires obtaining the detailed status of all nodes through communication, which costs considerable time and energy \cite{taskall}.
  
  Digital twin (DT) is an appealing technology that can assist UASNs in tackling the above issues. 
  By creating a high-fidelity digital replica of the actual UASN, DT reflects the real network states in real-time \cite{dt}. Moreover, DT can also implement control to the physical entities in real UASN to achieve a bidirectional closed-loop feedback process of dynamic information \cite{dta}.
  However, empowering existing UASNs with DT presents several challenges, summarized as follows:

  (a) \textit{Complexity of DT architecture.}
  Due to diverse devices, topologies, and protocols, UASN's performance is affected by the configurations of both node (such as power and sending rate) and network (such as routing tables and time slot allocation) \cite{UIOT}.
  However, UASN node and network differ in the domain, data amount, and controllable resources, thus DTs for them require different customized construction and control methods.
  A single-layer DT architecture, which simultaneously handles node and network in the same methods, struggles to address their diversity.
  Such an architecture fails to achieve accurate UASN replication, resulting in weak control of real UASN and poor maintainability of DT.

  (b) \textit{Consumption for constructing DT.}
  DT construction requires long-term communications to obtain massive real network data, which costs considerable time and energy, reducing the network lifetime \cite{DS}. Although low real-time data's collection can conserve resources, it decreases the DT fidelity.
  
  (c) \textit{Incompatibility of UASNs methods to DT.} 
  Compatibility of DT, UASN, and specific UASN studies is a necessary prerequisite for improving the UASN performance. However, existing UASN studies are designed only for real UASN and lack interactions with DT, failing to directly implement in DT-based architecture.

  To this end, we propose a Digital Twin-based intelligent Network Architecture (DTNA),
  which enhances UASN with strong environmental adaptability, intelligence, and multifunctionality. 
  We discuss the main contributions as follows.
  \begin{enumerate}
  
    \item {We propose a layered Digital Twin-based Network Architecture (DTNA) for UASNs, consisting of real network, local DT, and global DT. 
    Local DT is constructed by extracting the local data of each node in the real network. Global DT is aggregated from local DTs.
    The layered design aligns with UASNs' data flow characteristics, simplifying the DT construction and maintenance, and improving the flexibility of control over the real network.}

     \item {We develop a resource allocation paradigm in local DT (RAPD) to enhance the timeliness and robustness of existing resource allocation algorithms.
     RAPD realizes rapid perception across diverse performance metrics to detect performance variation. Moreover, by iteratively evaluating and optimizing allocation schemes, RAPD can prevent the UASN from employing deviant resources to transmit packets.}
  
    \item {We propose a collaborative MARL framework in global DT (CMFD) to accelerate AI model training for collaborative tasks. 
    In CMFD, agents decide strategies based solely on local observations without external information exchange and upload the decentralized transitions to global DT for training.
    CMFD can patch impaired data caused by node malfunction and communication faults. 
    In addition, CMFD supplements extensive DT data to increase training data diversity. 
    }
  
    \item {We propose a task-oriented network slicing in global DT (TNSD) to assist multi-task scheduling methods with realizing flexible scheduling. 
    TNSD unifies the decomposition and extraction of various task demands to adapt to heterogeneous application scenarios. Moreover, TNSD offers comprehensive network status in a cost-effective way, allowing task scheduling methods to rapidly obtain the detailed network status.
    }
  
  \end{enumerate}

\section{Related Work}

Due to the desire for UASNs with strong environmental adaptability, intelligence, and multifunctionality, extensive researches have been conducted on resource allocation, MARL algorithm design, and multi-task scheduling.

\subsection{Resource Allocation}
Existing underwater resource allocation studies are dedicated to appropriately allocate network and communication parameters.
In \cite{TDA} and \cite{AP}, nodes estimate propagation delays to obtian the channel occupancy status, enabling appropriate temporal channel allocation to reduce collisions.
\cite{DOTS} allows nodes to overhear the transmission schedules of neighbors to allocate time slots, realizing collision-free simultaneous transmission.
\cite{NOMA} designed a hybrid non-orthogonal multiple access-based Multiple Access Control (MAC) protocol to jointly allocate transmission power and codebook, which achieves efficient concurrent communication and optimizes energy consumption.
\cite{pcmac} allows nodes to allocate power based on the signal strength of Request to Send (RTS) packets during the handshake phase, reducing energy consumption while ensuring successful transmission.
\cite{Deep} employs Deep Reinforcement Learning (DRL) to allocate time slots, which maximizes the network throughput without prior knowledge of propagation delays and transmission strategies.

\subsection{MARL Algorithm Design}
MARL has been widely utilized for underwater collaborative tasks such as target tracking, data collection, and power management \cite{mao}.
\cite{mappo} proposed an energy-efficient underwater multiobjective scheduling scheme for real-time AUV target hunting and environment searching, enabling AUVs to autonomously make decisions and collaboratively complete the collaborative task through limited information interaction.
\cite{EDC} proposed a MADDPG-based multiple AUVs trajectory planning algorithm that considers underwater obstacles and currents to maximize the collection rate and energy efficiency.
\cite{ACH} presented a distributed MARL-based collaborative power allocation approach to allocate the transmission power in a
semi-cooperative way, achieving fair-effective communication and robustness.
\cite{AU} implements multi-agent generative adversarial imitation learning from expert demonstrated trajectories, achieving formation control and obstacle avoidance for multiple unmanned underwater vehicles.

\subsection{Multi-Task Scheduling}
With the increasing demand for underwater applications, UASNs need to serve multiple tasks simultaneously.	
Existing researches on task scheduling primarily concentrate on demand extraction and execution optimization for single-type tasks, with minimal emphasis on collaborative scheduling for multiple tasks.
\cite{Unified} proposed a unified framework to schedule emergency and non-emergency tasks in multi-AUV data collection, which dynamically integrates the importance of cases and the accesses cost from the regional perspective, improving the applicability in a wide range of scenarios.
\cite{Task} proposed a method for task-oriented intelligent networking demand extraction, which can provide personalized networking schemes according to the task type, satisfying the demand of different tasks on networking performance. 
\cite{UVMS} presented a task prioritization strategy based on a generic underwater task goal classification transformation, which divides the complex underwater operational tasks into generic task combinations and optimizes the resources consumption during the whole task.
\cite{taskac} presented a task allocation framework that can form an optimal AUV team for accomplishing critical missions, maximizing energy efficiency and guaranteeing mission completion within time constraint.

\vspace{1\baselineskip}

However, the practical implementation of the above studies in UASNs faces significantly serious challenges.
Existing resource allocation methods tend to ignore the spatial and temporal variability of the underwater environment.
The fixed allocation strategies relying on static information cannot handle dynamic changes such as node drift, channel variations, and external interference, thus requiring reallocation of resources.
Moreover, the packet loss due to high BER impairs the DRL-based methods' observation, resulting in inappropriate allocation schemes.
%
The practical deployment of MARL also faces serious challenges. 
Due to the long propagation delay, low transmission rate, and high BER, MARL model training in the real network encounters challenges such as long interaction times, data scarcity, and data impairment.
Considering the inefficiency of training data collection, existing underwater MARL algorithms rely on simulators for model training, which tend to be low confidence because their simulators often idealize the dynamic and severe acoustic communication environments.
%
In existing multi-task scheduling methods,
dividing the network into subnets for various tasks simplifies scheduling and reduces conflicts among tasks,
but rigidizes the network topology and limits scheduling flexibility.	
Hindered by the high latency and energy scarcity, centralized scheduling that requires obtaining the gloabl network status is not feasible.
In conclusion, the current UASNs' architecture suffers from unique underwater characteristics, failing to satisfy the timeliness, efficiency, and flexibility as the existing UASNs studies requested. 

\subsection{Digital Twin Technology}
To break the above bottlenecks, we introduce DT in UASNs, which is one of the enabling technologies in internet of things systems \cite{SDT} \cite{Ho}. In recent years, DT has been widely applied in various fields, such as intelligent driving, computation offloading, multi-Unmanned Aerial Vehicle (UAV) navigation.
\cite{DTV} leverages transfer learning to facilitate DT synchronization, and achieves optimal joint access network selection and power level allocation in DT-enabled internet of vehicles.
\cite{DTVL} proposed a DT-empowered Mobile Edge Computing (MEC) framework, which empowers edge nodes with intelligence to achieve safe, efficient, and intelligent Connected and Automated Vehicle (CAV) lane-changing. 
\cite{DeepF} proposed a DT-enabled DRL training framework where the DRL model learns from DT, solving the inefficiency of data collection faced by the real-world UAVs.

In these studies, DT provides high-fidelity replica for various internet applications, enabling global observation, low-cost trial and error, and trend prediction. Based on the above advantages, DT is promising to deal with UASN studies' constraints of untimeliness, inefficiency, and inflexibility. Therefore, we design the DTNA to assist UASNs' development.

\section{Digital Twin-based Network Architecture for UASNs}

\begin{figure*}[ht]
  \centerline{\includegraphics[width=0.97\textwidth]{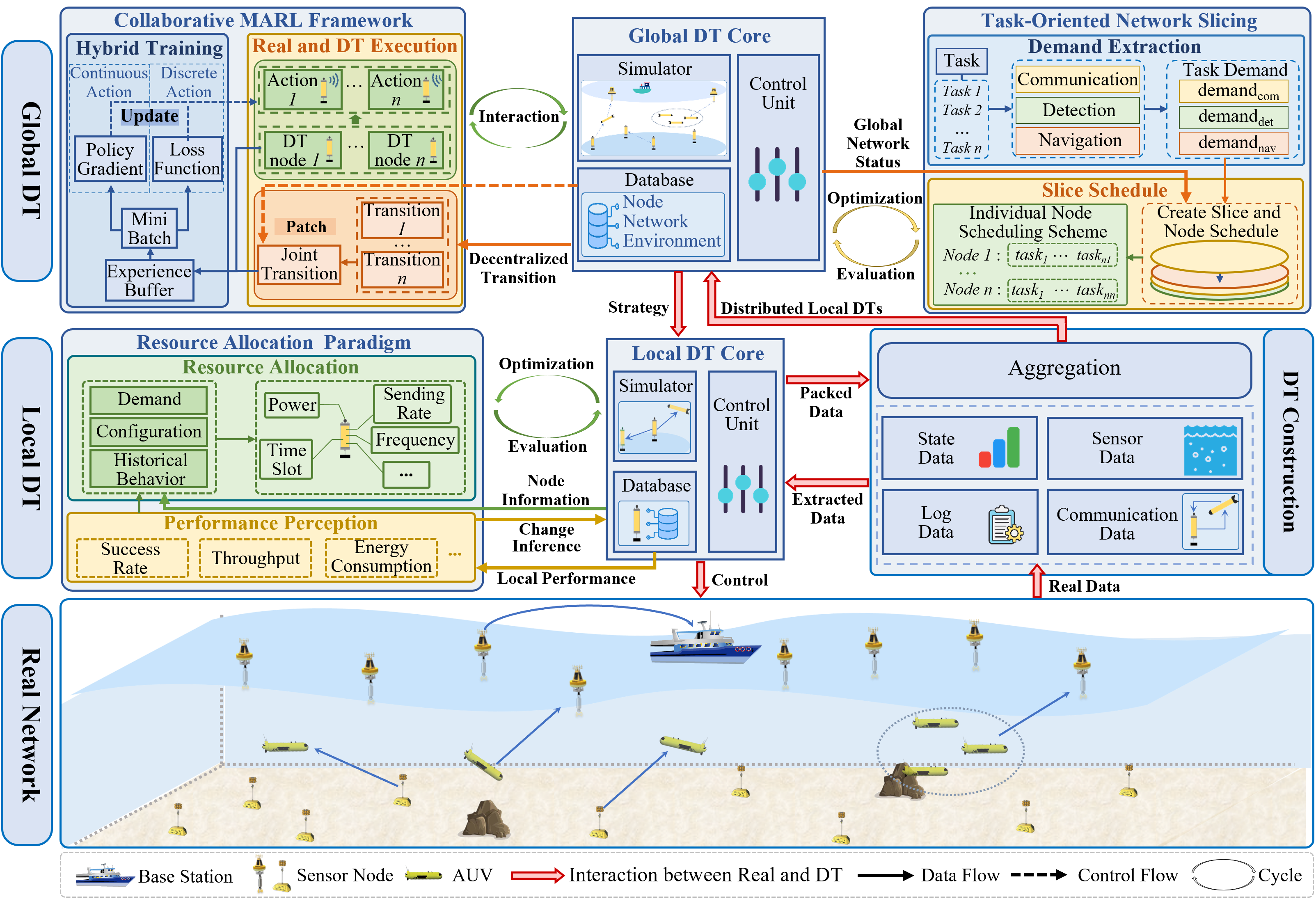}}
  \caption{Framework of DTNA. DTNA consists of real network, local DT, and global DT. Local DT extracts the local data to construct the digital replica of the node's surroundings. Global DT is aggregated from local DTs, representing the overall network. 
  We propose DT-empowered approaches to enhance the existing resource allocation, MARL, and multi-task scheduling methods, improving the environmental adaptability, intelligence, and multifunctionality of UASNs.}
  \label{fig1}
\end{figure*}

In this section,
we propose DTNA to address the various limitations encountered by current UASNs.
As shown in Fig.\ref{fig1}, our proposed DTNA integrates three layers: real network, local DT, and global DT.
\subsection{Real Network}
The real network of DTNA comprises a set of surface nodes and underwater nodes.
In the real network,
all nodes are equipped with acoustic modems to communicate with each other for transmitting data packets or issuing commands.
Nodes with sensors can collect the surrounding environmental information.
Moreover, the mobile nodes are capable of mobile data collection, communication relaying, and network extension and recovery.

Based on the above description, the nodes in the real network generate various types of data during their lifetime:
\begin{itemize}
 
  \item{\textbf{State data} refers to the node's local transmission configuration and device status, such as transmission power and residual energy. For a mobile node, it also includes movement parameters such as speed and attitude.}
  \item {\textbf{Sensor data} refers to the local data obtained by various sensors equipped on the node, including temperature, conductivity, obstacles' location, current velocity, etc.}
  \item {\textbf{Communication data} reflects the transmission information within communication range, such as packet flow, propagation delay, signal-to-noise ratio, packet size, etc.}
  \item {\textbf{Log data} refers to the execution logs of the node's intelligent methods at runtime, like the transitions of MARL algorithms.
   }
\end{itemize}
\subsection{Real and DT Interaction}
The aforementioned real data is automatically uploaded to construct the local DT in the following way.
State and sensor data are immediately uploaded whenever they are changed or updated during operation.
Moreover, the nodes parse the communication data of the packets transmitted by neighbors and then uploads it.
In addition,
the intelligent methods equipped with real-time interfaces upload execution log data after making decisions.

All the uploaded real data is extracted to construct the local DT.
Local DT construction involves no additional data exchange, with all operations performed locally. 
In contrast, global DT construction includes \textit{ aggregation} of all the packaged DT data from each local DT, in addition to collecting and extracting local real data.


At the network runtime,
global DT distributes the produced intelligent policy models, task scheduling schemes, and other strategies to the local DTs for execution.
Local DT executes the corresponding strategies based on the current DT status and implements control over the real network.

The specific details of local DT and global DT are presented in the following.

\subsection{Local DT}
Local DT of DTNA is lightweight and low-power, deployed on each node with computing capabilities.
Local DT leverages monitoring and simulation capabilities to provide resource allocation algorithms with rapid performance perception and iterative optimization.
Local DT realizes the digital replica of the node's surroundings,
which is represented as the node's local, neighboring, and environmental information.
\begin{itemize}
  \item {\textbf{Local information} is extracted from state, communication, and log data, including node configuration, transmission parameter, and historical behavior. Local information is responsible for providing node state monitoring and historical behavior records. Moreover, it assists in trend prediction and method optimization.}
  \item {\textbf{Neighboring information} is extracted from sensor, communication and log data, including the neighboring nodes' transmission parameters, location, and historical behavior. Neighboring information provides performance evaluation and trend prediction with the neighboring behavior, link status, and network topology.}
  \item {\textbf{Environmental information} is extracted from sensor and communication data, including obstacle locations, channel quality, currents, and so on. Environmental information provides the physical and acoustic characteristics of the surrounding environment to simulate the impact on the node's behavior.}
\end{itemize}

\begin{figure*}[ht]
  \centerline{\includegraphics[width=0.8\textwidth]{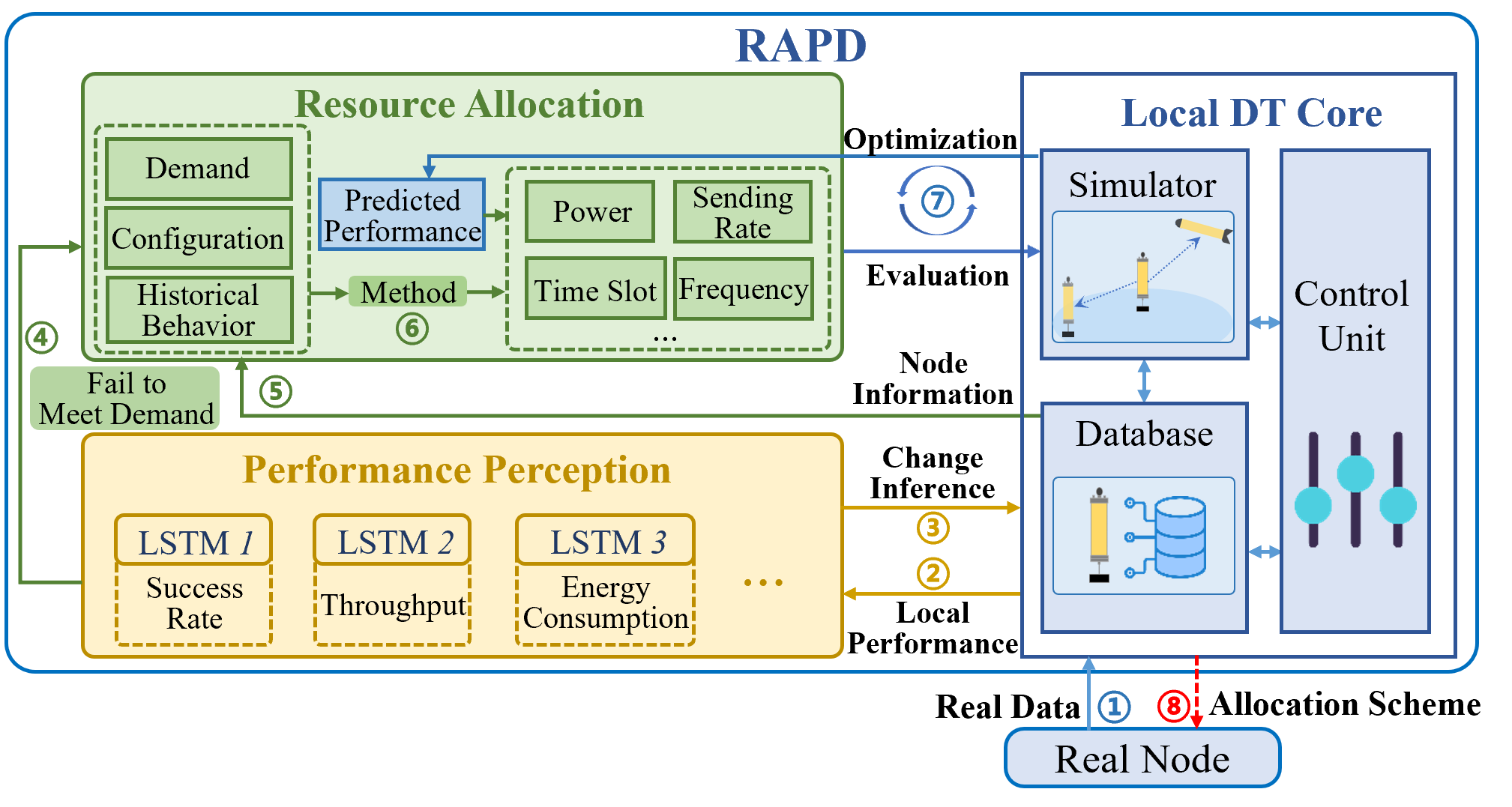}}
  \caption{Flowchart of RAPD. RAPD consists of local DT core, performance perception, and resource allocation. RAPD leverages the high-fidelity replica to rapidly perceive performance variations and iteratively optimize allocation scheme when performance metrics fail to meet demand. 
  }

  \label{fig2}
\end{figure*}
\subsection{Global DT}

Global DT of DTNA is deployed on the central nodes,
which are typically the sink node in data collection or the leader node of mobile clusters.
Because such nodes have sufficient energy and receive more data than other relay nodes, allowing for a comprehensive network knowledge.
Global DT constructs a digital replica, capturing each node's state and behavior, network's topology and configuration, and environmental characteristics.
Global DT patches real data and provides extensive DT data to accelerate model training for collaborative tasks.
Moreover, it extracts task demands as unified and rapidly provides a comprehensive network status at a low cost for multi-task scheduling methods.

In DT \textit{aggregation},
each node uploads the local DT data to the central nodes. 
We design specific rules for different types of nodes to upload data due to their diversity in communication and energy.
To avoid costly battery replacement, fixed sensor nodes piggyback DT data on normal data packets, rather than sending actively.
In addition to the piggybacking way, sufficient-energy and high-performance nodes, like AUVs, actively upload local DT parameters based on the pre-configurations or commands.
The \textit{local}, \textit{neighboring}, and \textit{environmental} information in local DT are aggregated and organized into \textit{node}, \textit{network}, and \textit{environmental} information in global DT, enabling it to construct and maintain the replica of the gloabl network.

Maintaining the global DT does not solely rely on successful communication. 
The real network scales represented by each local DT overlap to some extent. 
Even if a local DT cannot upload data, the absent network portion is potentially supplemented by other local DTs.
In addition, global DT can utilize historical information to predict the status of network areas that lack DT data.
Therefore global DT achieves high-fidelity and robust replica of the real network.

\section{Resource Allocation Paradigm\\ based on Local DT}

In this section, we propose RAPD to enhance the environmental adaptability of resource allocation algorithms.
As shown in Fig.\ref{fig2}, RAPD comprises three parts: local DT core, performance perception, and resource allocation.

\subsection{Local DT Core for RAPD}

Local DT core represents a digital replica of the node's surroundings, consisting of three components: control unit, database, and simulator.

\textbf{Control unit} is responsible for data processing, simulator drivering, and scheme deployment. 
Control unit analyzes various extracted data from DT construction to obtain local, neighboring, and environmental information, which are stored in the database and updated based on the subsequent real-time data.
In addition, control unit drives the simulator to provide performance evaluation or generate DT data as required by the intelligent methods in DT.
Control unit can also assess whether the intelligent methods' schemes meet the demand and deploy them to the real network for execution.

\textbf{Database} stores the node's local, neighboring, and environmental information, allowing the real-time monitoring of the node's surroundings.
It can provides real-time performance metrics based on the historical behavior in local information.

\textbf{Simulator} is the most critical part of the DT core. 
Simulator obtains the network parameters from the database to conduct performance evaluation or DT data supplementation as required by the control unit.
Moreover, it can be dynamically configured to simulate unexpected situations to enhance the robustness of the intelligent methods.
Simulator is capable of generating a variety of performance metrics to provide optimization proposals for the control over the real network.
With the above features of the simulator, the local DT core is able to control and optimize the real network based on the high-fidelity replica, achieving the bidirectional influence between real and DT.


In RAPD, local DT core is able to provide the performance metrics for performance perception and the real-time node information for resource allocation.
Local DT core can also evaluate the allocation schemes over the current network.

\subsection{Performance Perception}

To rapidly perceive performance degradation, we develop performance perception based on Long Short-Term Memory (LSTM) \cite{LSTM}. Moreover, RAPD can infer the specifics of external changes and internal failures based on LSTM and the local DT core.

Local DT core continuously collects various real-time performance metrics based on local and neighboring information, such as data success rate, throughput, energy consumption, and so on. 
Such performance metrics over a historical period is input to the corresponding LSTMs to obtain the predicted performance expectations.


The LSTM is widely used for prediction, in which the LSTM cell is capable of storing historical sequential data and analyzing the temporal correlation.
Each LSTM for performance perception is composed of several LSTM cells in series. LSTMs take the historical performance metrics \( [W_{k-n,d}, W_{k-n+1,d}, \dots, W_{k-1,d}] \) as input, and generate the expected performance metrics \( W_{k,d}^v \).

If the real performance metrics fall below the expectations by more than a certain threshold, it indicates internal failures or external interference beyond the current local DT coverage, which lead to a performance degradation.
Internal failures can be rapidly distinguished because the local DT core monitors the real-time node state, while external changes require to be inferred by the local DT core.
Based on the difference between the real performance metrics and expectations from LSTM, the local DT core simulates various external changes to evaluate the performance of the current allocation scheme, aiming for a high fitting between evaluated and real performance metrics to accurately infer external changes.

Finally, resource allocation is initiated to adjust network parameters to meet the demand.

\subsection{Resource Allocation}

RAPD does not define the specifics of the resource allocation algorithms, but provides a generalized model.
Resource allocation makes preliminary schemes based exclusively on local information, and the schemes are iteratively evaluated and optimized by the DT core to obtain optimal results. The detailed process of resource allocation is described below.

Firstly, local DT core delivers real-time node information to the resource allocation as input, including demand, configuration, and historical behavior.
The demand refers to the minimum thresholds for performance metrics such as throughput, latency, loss, and energy consumption, which are derived from the performance expectations and potential commands.
The configuration indicates the node's local parameters such as power, bandwidth, and sending rate. It also contains neighboring information obtained from the particular observation.
The historical behavior refers to the node's local transmission or movements over a historical period, as well as the observed neighboring behavior.

\begin{figure*}[ht]
  \centerline{\includegraphics[width=0.85\textwidth]{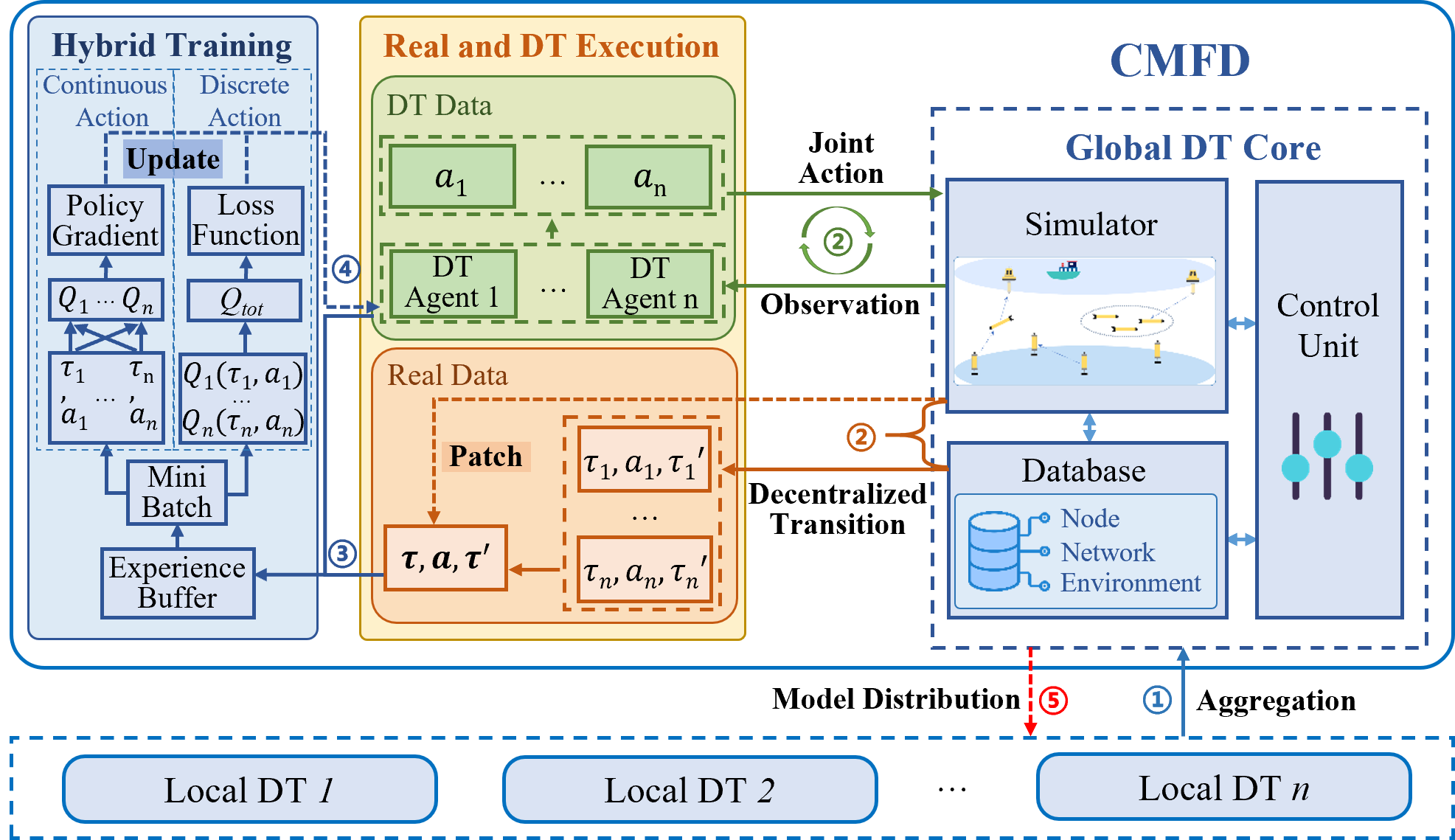}}
    \caption{Flowchart of CMFD. CMFD consists of global DT core, real and DT execution, and hybrid training. 
    CMFD aggregates and patches scarce real data, and generates extensive DT data through global DT core's simulator.
    Finally, CMFD performs hybrid training based on such both types of data and distributes models to each local DT.}
  \label{CMFD}
\end{figure*}

Then the resource allocation algorithm generates preliminary allocation scheme based on the available inputs from local DT core, which includes the node's parameters, such as power, time slots, and frequency.
The preliminary scheme is delivered to the latest local DT core to conduct a spatial and temporal evaluation to obtain the predicted performance metrics.
Next local DT iteratively fine-tunes and evaluates the scheme parameters to obtain the solution with best performance in the replication of the current local DT core.

Finally, local DT core deploy the optimal resource allocation scheme to the real network for execution.

\section{Collaborative MARL Framework\\ based on Global DT}

In this section, we propose CMFD to accelerate AI model training for collaborative tasks.
We do not qualify specific algorithm details, but propose an DT-based execution and training paradigm for collaborative tasks with discrete and continuous action spaces.
As shown in Fig.\ref{CMFD}, CMFD comprises three components: global DT core, real and DT execution, and centralized hybrid training.



\subsection{ Global DT Core for CMFD}

Global DT core in CMFD offers a digital replica of the overall network. 
Same as local DT core, global DT core consists of three components: control unit, database, and simulator. 
Each component serves the same function as its counterpart in local DT core, with additional features.
The control unit of global DT processes aggregated data and distributes AI models.
The database differs slightly from that of local DT, 
which stores the \textit{node}, \textit{network}, and \textit{environmental} information aggregated from the \textit{local}, \textit{neighboring}, and \textit{environmental} information of each local DT.

In CMFD, global DT core is required to aggregate and patch the decentralized transitions.
It also provides a high-fidelity network replica as an alternative interactive environment to rapidly generate extensive DT data in a low-cost way.
Global DT core can extend network scenarios to simulate contingencies such as node malfunction and external interference,
increasing the diversity of training data and enhancing the AI model's adaptability to unexpected situations.

\subsection{ Real and DT Execution}

Real and DT execution refers to MARL algorithms interacting with real or simulated environments to make autonomous decisions and execute actions. It generates data for centralized model training.

For the mentioned underwater collaborative tasks such as target tracking and data collection, we model them as a Decentralized Partially Observable Markov Decision Process (Dec-POMDP),
which is given by a tuple $G=\langle \mathcal{N} ,\mathcal{S} ,\mathcal{A} ,P,\mathcal{R},\mathcal{O}  ,O,\gamma \rangle $.
Here $\mathcal{N} \equiv\{1, . . . , n\}$ denotes the finite set of agents and $s \in S$ describes the true state of the environment.
At each time step, each agent $i \in \mathcal{N} $ selects a discrete or continuous action $a_i \in \mathcal{A} $, forming a joint action $ \bm{a} \in \bm{\mathcal{A}} \equiv \mathcal{A}^n$.
This results in a transition to the next state $s'$ according to the state transition function $P(s'|s, \bm{a}):\mathcal{S}\times \bm{\mathcal{A}}\times \mathcal{S}\rightarrow [0,1]$ and a team reward $r=\mathcal{R}(s, \bm{a})$.
Due to the partial observability, each agent $i \in \mathcal{N} $ draws an individual partial observation $\tau_i \in \mathcal{O}$ from the observation kernel $O(s, a)$.
$\gamma \in [0, 1)$ is the discount factor, which indicates the long-term effect of the current action.

\begin{figure*}[ht]
  \centering
  \captionsetup[subfigure]{font=small} 
  \subfloat[Discrete Action]{
      \includegraphics[scale=0.58]{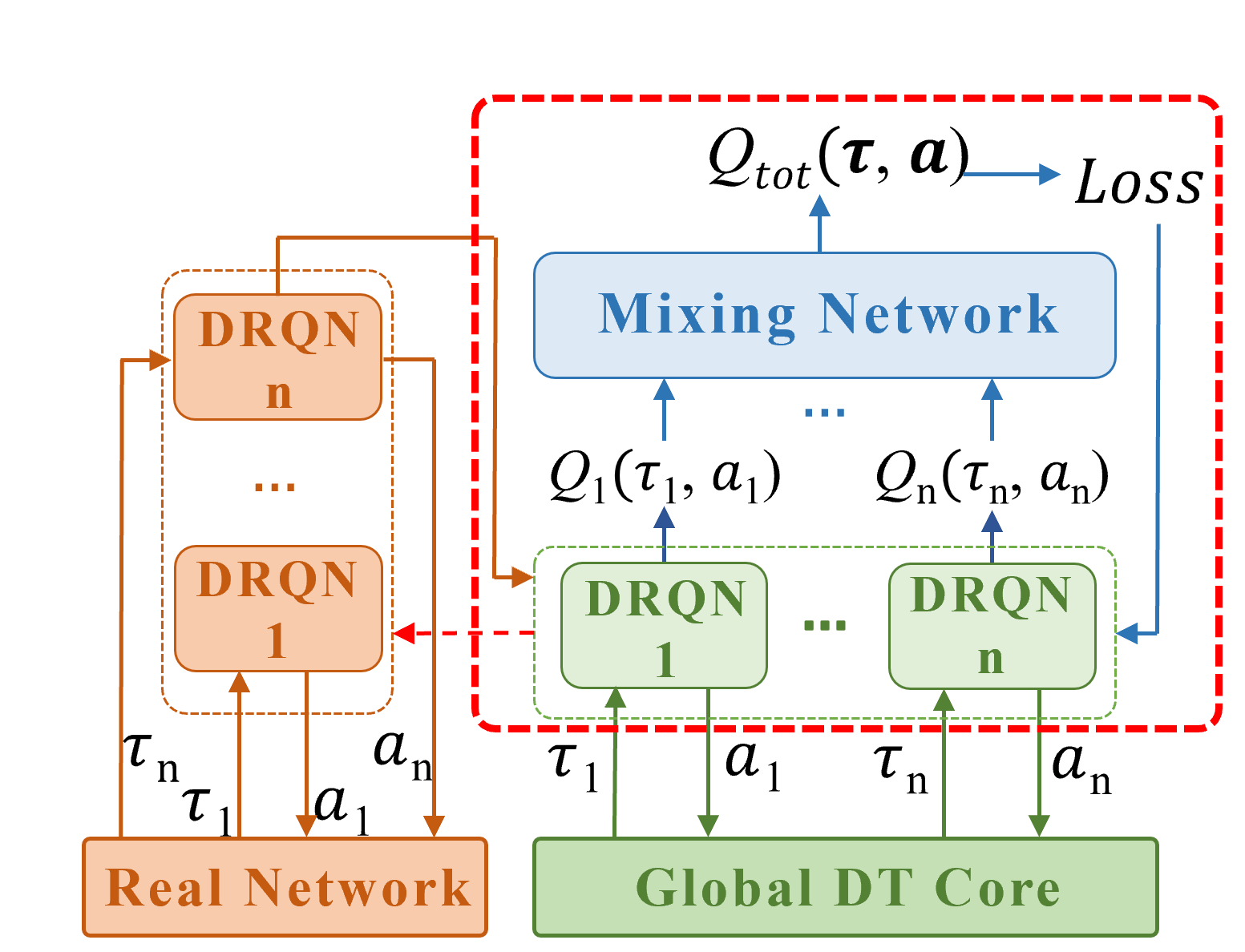}}
  \subfloat[Continuous Action]{
      \includegraphics[scale=0.58]{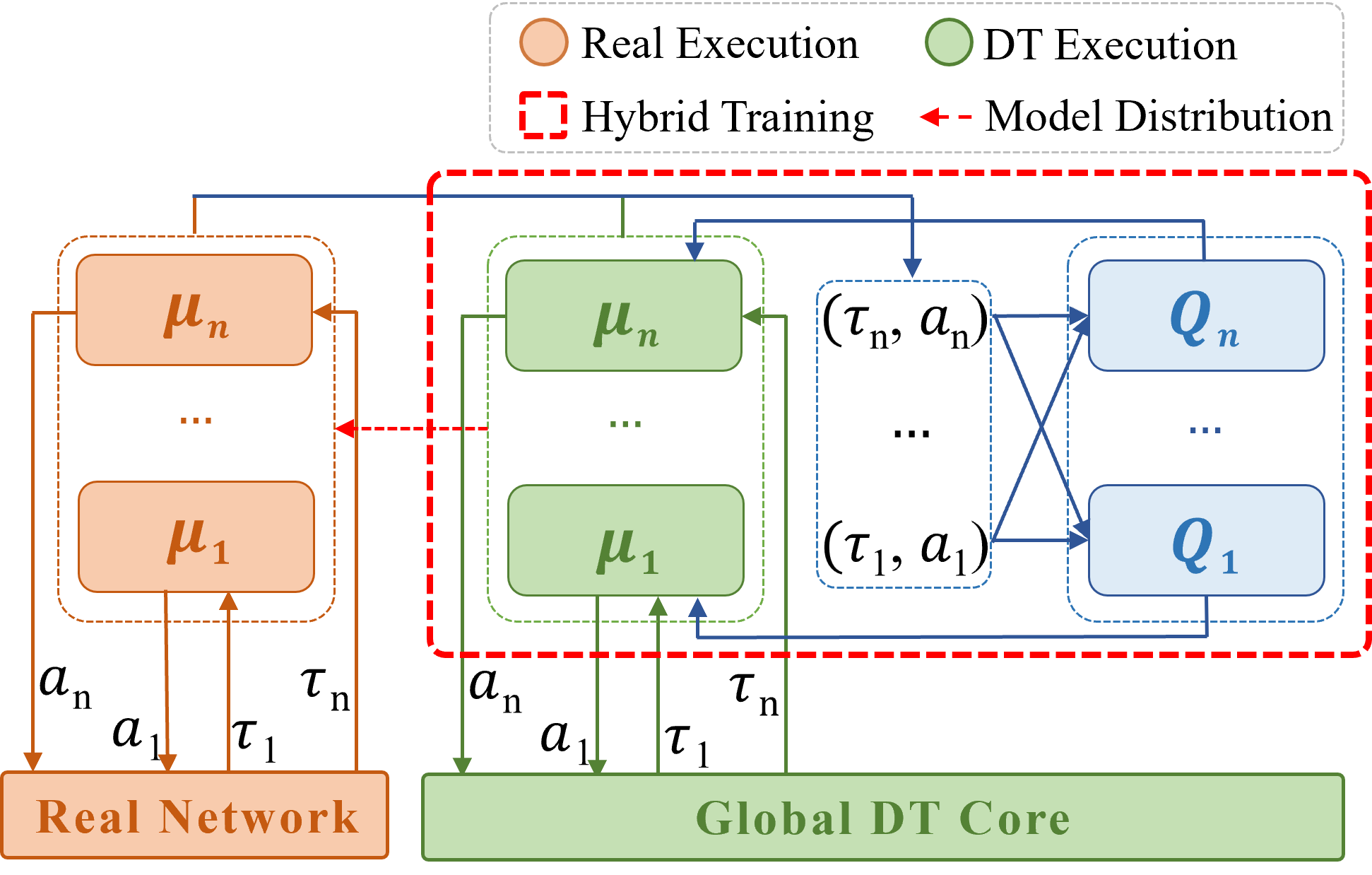}}
  \caption{Execution and training paradigm of MADRL models with discrete or continuous action space in CMFD.}
  \label{2a}
\end{figure*}

As shown in Fig.\ref{2a}, 
each agent with \textbf{discrete} action space has a Deep Recurrent Q-network (DRQN), which receives the current local observation as input at each step and estimates its individual Q-value function $Q_i(\tau_i, a_i; \beta_i)$. Then agent chooses the action corresponding to the maximum Q-value. 
Each agent with \textbf{continuous} action space has a deterministic policy network $\mu_i(\tau_i; \beta_i)$,
which receives the current local observation as input at each step and outputs a deterministic action.


Decentralized agents in local DTs do not have model update capability. 
To trade off exploration and exploitation, agents randomly choose actions with a probability of $\varepsilon$, and choose actions guided by models with a probability of $1-\varepsilon$.
Each model in agent $i$ chooses action $a_i$ individually based on only local observation $\tau_i$ and gets next observation $\tau_i^{\prime}$. \textit{Real transition} $\{\tau_i,a_i,\tau_i^{\prime}\}$ is uploaded voluntarily to global DT in the initial training stage, but gradually evolves to be piggybacked by the normal data packets.

The decentralized transitions of all individually executed agents for each step are aggregated in global DT to generate joint transition $\{\bm{\tau},\bm{a},\bm{\tau^{\prime}}\}$, where $\bm{\tau} = \{\tau_i|i \in N \}$ is the joint observation, $\bm{a} = \{a_i|i \in N \}$ is the joint action of all agents, and $\bm{\tau^{\prime}} = \{\tau_i^{\prime}|i \in N \}$ is the next joint observation. 
$\bm{\tau}$ and $\bm{\tau^{\prime}}$ in the joint transition may be impaired due to node failures, packet errors, and collisions.
Global DT core patches the impaired transitions by utilizing the high-fidelity replica of the real work.
Finally, the real joint transitions are stored into the experience buffer.


Global DT in CMFD contains identical models with the same parameters as the decentralized agents. 
In DT execution, agents interact with global DT core's simulator to generate DT transitions. 
Each \textit{DT transition} $\{\tau_i,a_i,\tau_i^{\prime}\}$ is aggregated into joint transition $\{\bm{\tau},\bm{a},\bm{\tau^{\prime}}\}$ and stored into the experience buffer.
Real and DT transitions are stored indiscriminately in the experience replay buffer.

\subsection{ Centralized Hybrid Training}

CMFD conducts centralized hybrid training in global DT, employing a hybrid of real and DT data to efficiently update each agent's model. 
We propose two training approaches for models with discrete and continuous action space.

\noindent\textbf{1) Discrete Action Space}

In the centralized hybrid training for models with discrete action space, we import a mixing network to aggregate individual Q-values from real or DT execution \cite{qmix}.
CMFD maximizes the aggregated Q-value while guaranteeing monotonicity of the aggregated and individual values, improving the global performance and interagent collaboration.

As shown in Fig.\ref{2a}(a), the hybrid training consists of the mixing network and all DT DRQNs.
In the training process, CMFD samples a batch of transitions $\{\bm{\tau},\bm{a},\bm{\tau^{\prime}}\}$ from experience buffer and obtain all $Q_i(\tau_i,a_i)$ using DT DRQNs. 
The mixing network $f(\cdot)$ makes a mixture of all $Q_i(\tau_i,a_i)$ to produce the joint action-value function $Q_{tot}$ as

\begin{equation}
  \label{Qtot}
  Q_{tot}(\boldsymbol{\tau}, \boldsymbol{a}) = f(Q_1(\tau_1,a_1),\dots,Q_n(\tau_n,a_n)).
\end{equation}

$Q_{tot}$ has the same monotonicity with the single agent value function as

\begin{equation}
  \label{arg}
\arg \max _{\boldsymbol{a}} Q_{\text {tot}}(\boldsymbol{\tau}, \boldsymbol{a})=\left(\begin{array}{c}
  \arg \max _{a_{1}} Q_{1}\left(\tau_{1}, a_{1}\right) \\
  \ldots \\
  \arg \max _{a_{n}} Q_{n}\left(\tau_{n}, a_{n}\right)
  \end{array}\right).
\end{equation}

The monotonicity can be enforced by imposing a constraint
on the relationship between $Q_{tot}$ and each $Q_{i}$ as

\begin{equation}
  \label{mon}
  \frac{\partial Q_{\text {tot}}(\boldsymbol{\tau}, \boldsymbol{a})}{\partial Q_{n}({\tau}_{i}, a_{i})} \geq 0, \quad \forall n \in\{1,2, \ldots, N\}.
\end{equation}

To guarantee (\ref{mon}), we use a hypernetwork to generate nonnegative parameters as the weights $W_i$ of the mixing network.
The hypernetwork takes the global network state $s$ as input, which is inferred from $\boldsymbol{\tau}$ by global DT core.
Biases $b_i$ of the mixing network are generated similarly, but can be negative.

The DRQN agents in global DT are trained end-to-end to minimize the following loss

\begin{equation}
  \label{loss}
Loss(\theta)=\frac{1}{b}\sum_{j=1}^{b}\left(y^{j}_{\text {tot }}-Q_{\text {tot }}\left(\boldsymbol{\tau}^{j}, \boldsymbol{a}^{j} ; \theta\right)\right)^{2},
\end{equation}

\begin{equation}
  \label{y}
y_{t o t}=r+\gamma \max _{\mathbf{a}^{\prime}} Q_{t o t}\left(\boldsymbol{\tau}^{\prime}, \mathbf{a}^{\prime}; \theta^{-}\right),
\end{equation}
where $b$ is the batch size, $\gamma$ is the discount factor, $\theta$ is the parameters of the evaluation network, $\theta^{-}$ is the parameters of the target network. $\theta$ of the evaluation network is updated by minimizing (\ref{loss}).
$\theta$ is assigned to $\theta^{-}$ every certain episodes, otherwise it stays unchanged between individual updates.

\noindent\textbf{2) Continuous Action Space}

In the centralized hybrid training for models with continuous action space, we import value networks for each policy network to incorporate information from other agents, allowing for a better understanding of the others' behavior and thus achieving better collaboration \cite{maddpg}.

As shown in Fig.\ref{2a}(b),
the hybrid training consists of all DT policy networks and the corresponding value networks.
In the training process,
CMFD samples a batch of transitions $\{\bm{\tau},\bm{a},\bm{\tau^{\prime}}\}$ from experience buffer. The centralized value network $Q_i$ is trained to minimize the following loss


\algrenewcommand\algorithmicindent{0.7em}  
\begin{algorithm}
  \caption{Centralized hybrid training}
  \label{train}
  \begin{algorithmic}[1] 
    \State Get the experience reply buffer $D$;
    \State Set the learning rate $\alpha$; set the discount factor $\gamma$;
    \For{$\textit{train}$ $\leq$ $\textit{train step max}$ in each $\textit{epoch}$}
    \State Sample a batch of hybrid DT and real transitions;
    \If {$\textit{discrete action space}$} 
    \State Get each DT agent's DRQN $\theta $ and target DRQN $\theta^{-} $;
    \State Update the DRQN by minimizing the loss as (\ref{loss});
    \If {$\textit{update-interval}$ train}
    \State $\theta^{-} = \theta$;
    \EndIf
    \EndIf
    \If {$\textit{continuous action space}$} 
    \State Get value network $\theta$ and target value network $\theta^{-} $ from Global DT;
    \State Get each DT agent's policy network $\phi$ and target policy network $\phi^{-} $;
    \State Update value network by minimizing the loss as (\ref{closs});
    \State Update policy network using the policy gradient as (\ref{pg});
    \If {$\textit{policy-update-interval}$ train}
    \State $\phi^{-} = \phi$;
    \EndIf
    \If {$\textit{value-update-interval}$ train}
    \State $\theta^{-} = \theta$;
    \EndIf
    \EndIf
    \EndFor
  \end{algorithmic}
  \end{algorithm}

\begin{equation}
  \label{closs}
Loss\left(\theta_i\right)= \frac{1}{b}\sum_j^b\left(y^j-Q_i^{\boldsymbol{\mu}}\left(\bm{\tau}^j, a_1^j, \ldots, a_N^j\right)\right)^2,
\end{equation}

\begin{equation}
  \label{cy}
  y^{j}=r_{i}^{j}+\left.\gamma Q_{i}^{\boldsymbol{\mu}^{\prime}}\left(\bm{\tau}^{\prime j}, a_{1}^{\prime}, \ldots, a_{N}^{\prime}\right)\right|_{a_{k}^{\prime}=\boldsymbol{\mu}_{k}^{\prime}\left(\tau_{k}^{j}\right)},
\end{equation}
where $\boldsymbol{\mu}=\left\{\mu_{1}\left(\tau_{1} ; \phi_{1}\right), \ldots, \mu_{n}\left(\tau_{n} ; \phi_{n}\right)\right\}$ is the set of all agents' current policy networks with parameter $\phi_{i}$, $b$ is the batch size, $\gamma$ is the discount factor. $Q_i^{\boldsymbol{\mu}}\left(\bm{\tau}^j, a_1^j, \ldots, a_N^j\right)$
is a centralized action-value function that takes the joint action $\bm{a} = \{a_i|i \in N \}$ and the joint observation $\bm{\tau} = \{\tau_i|i \in N \}$ as input, and outputs the Q-value for agent $i$.

The policy network of each agent $i$ in global DT can be updated by calculating the policy gradient as

\begin{equation}
  \begin{split}
    \nabla_{\phi_i} J = \frac{1}{b} \sum_j^b & \nabla_{\phi_i} \boldsymbol{\mu}_i\left(o_i^j\right) \\
    & \nabla_{a_i} Q_i^\mu\left(\bm{\tau}^j, a_1^j, \ldots, a_N^j\right)\Big|_{a_i=\boldsymbol{\mu}_i\left(\tau_i^j\right)},
  \end{split}
  \label{pg}
\end{equation}

\begin{equation}
  \label{aupdate}
  \phi_i = \phi_i + \alpha\bigtriangledown_{\phi_i} J(\boldsymbol{\pi}),
\end{equation}
where $\boldsymbol{\mu}=\left\{\mu_{1}\left(\tau_{1} ; \phi_{1}\right), \ldots, \mu_{n}\left(\tau_{n} ; \phi_{n}\right)\right\}$ is the set of all agents' current actor network with parameter $\phi_{i}$.
$\phi_i$ of the actor network is updated according to (\ref{aupdate}).
\begin{figure}[t]
  \centerline{\includegraphics[width=0.5\textwidth]{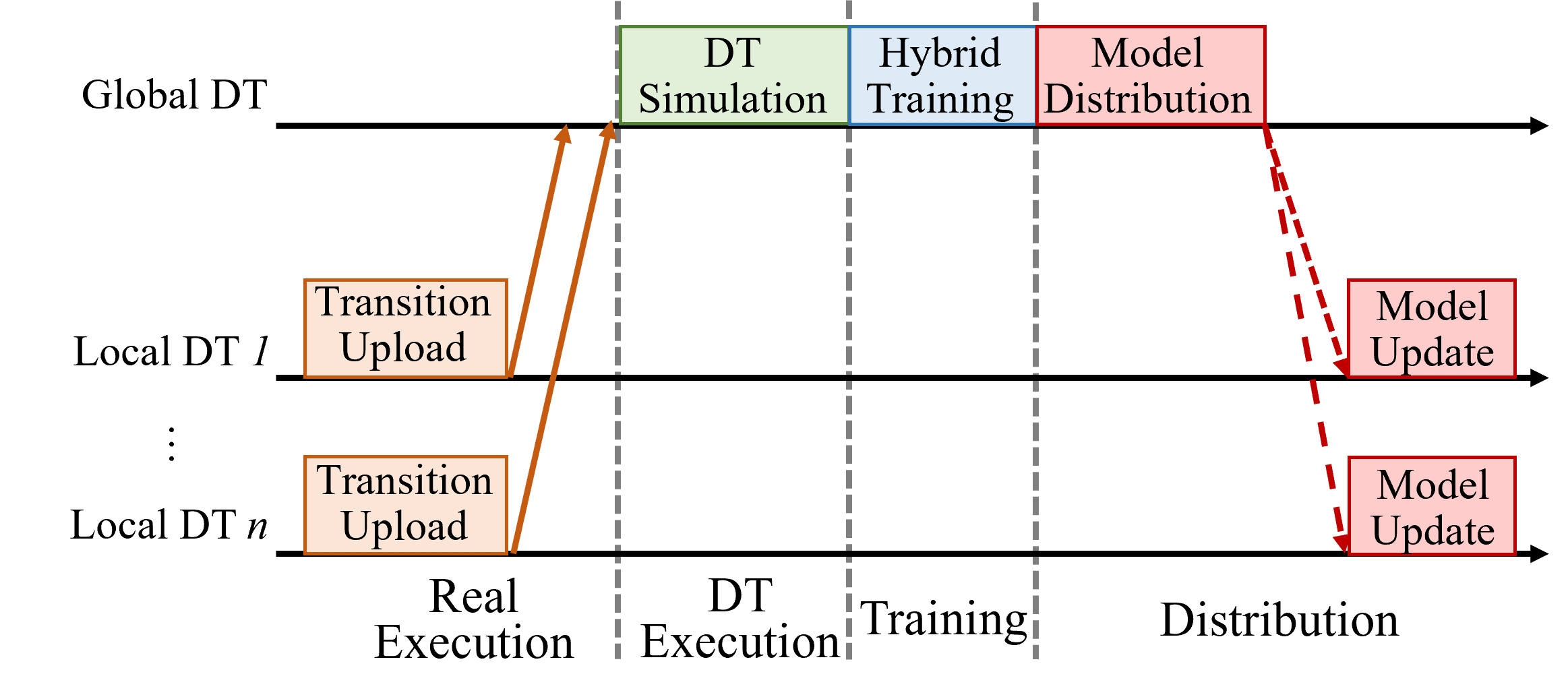}}
  \vspace{-10pt} 
  \caption{Timeline of CMFD execution stages.}
  \label{fig6}
\end{figure}

The centralized hybrid training of CMFD is detailed in Algorithm.\ref{train}.
Gloabl DT core can evaluate whether the centralized hybrid training is converging.
After obtaining the optimal collaborative strategy, global DT distributes the model parameters to the local DTs in each decentralized node.
The complete execution stages of CMFD are shown in Fig.\ref{fig6}.

\begin{figure*}[t]
  \centerline{\includegraphics[width=0.85\textwidth]{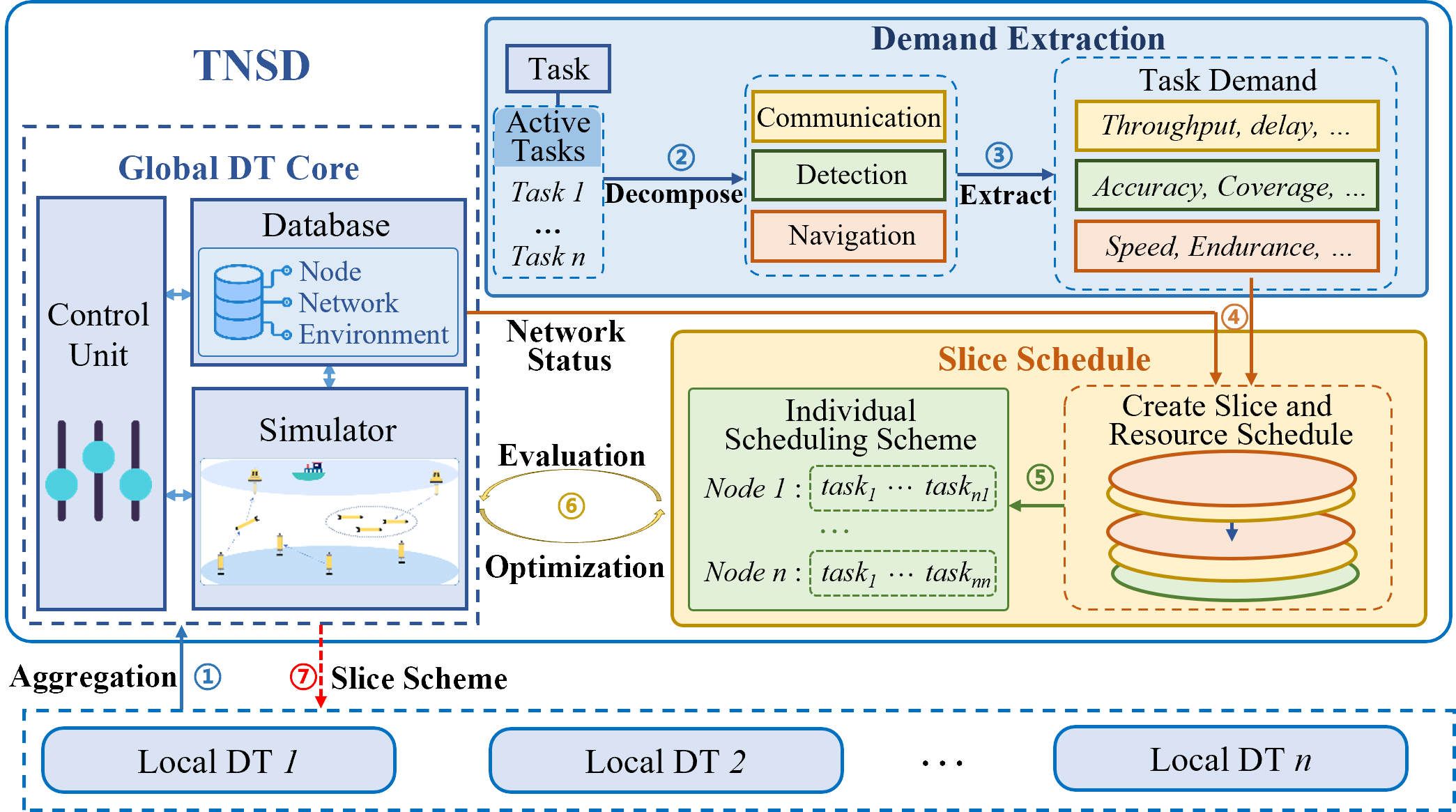}}
    \caption{Flowchart of TNSD. 
    TNSD unifies heterogeneous task demand extraction and rapidly provides a comprehensive network status in a cost-effective way.
    TNSD creates slice and schedules resources accordingly, and then the network slices are converted into individual node scheduling schemes, which are distributed to each local DT for execution.}
  \label{fig7}
\end{figure*}

\vspace{=1.5cm} 

\section{Task-Oriented Network Slicing\\ based on Global DT }
In this section, we propose TNSD to improve the flexibility of multi-task scheduling algorithms.
As shown in Fig.\ref{fig7}, TNSD comprises three components: global DT core, demand extraction, and slice schedule.

\subsection{Global DT Core for TNSD}

TNSD and CMFD share a common global DT core, but each emphasizes different functionalities.
In addition to aggregating each local DT to construct a digital network replica, the control unit of TNSD also consistently monitors the timeliness of the network scales represented by each local DT.
The control unit drives the simulator to predict events in network scales that lack local DT data, such as node drift, energy exhaustion, mobile node joining.

In TNSD, global DT core is required to provide the current global network status and predict network trends.
In addition, global DT core can evaluate the slice schemes spatially and temporally to obtain the corresponding performance for scheme optimization.

\subsection{Demand Extraction}

Demand extraction is responsible for task decomposition and demand extraction. 
Heterogeneous tasks differ significantly in emphasis on communication, detection, and navigation, which have diverse resource demands.
To unify and simplify task demand extraction, we decompose heterogeneous tasks into three subtasks: communication, detection, and navigation.

\begin{figure}[t]
  \centerline{\includegraphics[width=0.5\textwidth]{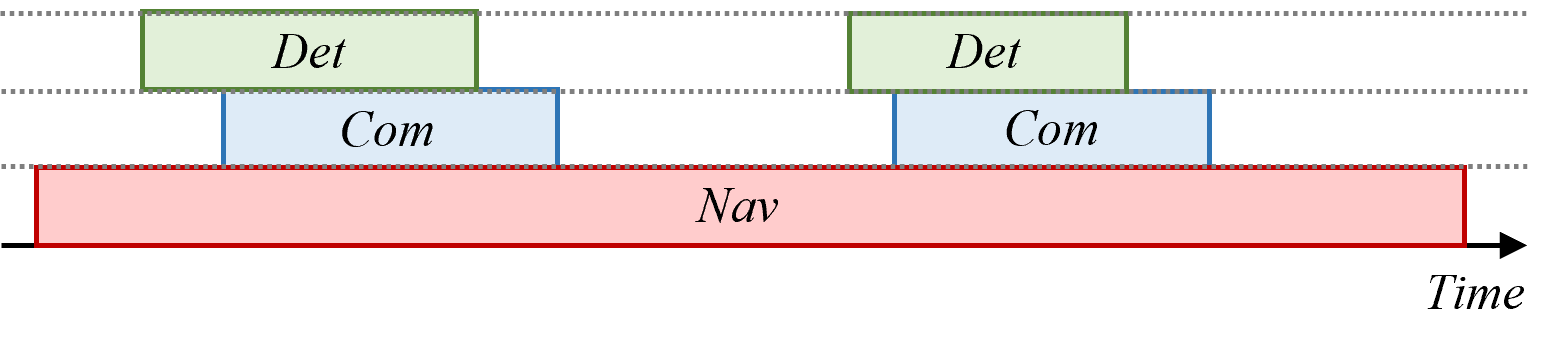}}
  \caption{The scheduling timeline $ST$ of a cruise task. Nodes navigate on a specified path (\textit{Nav}), detect in alert areas (\textit{Det}), and upload detection data (\textit{Com}).}
  \label{subt}
\end{figure}
The new task is delivered to TNSD for decomposition and then converted into a raw task matrix $RT = [rt_{com}, rt_{det}, rt_{nav}, ST] $ to abstract the task characteristics. 
Each type of subtask corresponds to a specific $rt$ as
\begin{equation}
\begin{aligned}
  \label{raw}
  & rt_{com} = [ source, destination, size, number,\dots], \\
  & rt_{det} = [ location, scale, duration, urgency, \dots ], \\
  & rt_{nav}  = [ destination, pathway, timing, \dots],
\end{aligned}
\end{equation}
where each $rt$ contains the essential parameters of the subtask to support the demand extraction.
$ST$ is the scheduling timeline of the task, which indicates the scheduling of the three subtasks over the entire task lifetime.
$ST$ is represented as
\begin{equation}
  \label{st}
  ST = \left[\begin{array}{c}
    st_{com} \\
    st_{det} \\
    st_{nav}
    \end{array}\right] = \left[\begin{array}{cccc}
    \mathit{[s_1,e_1]},  \cdots, \mathit{[s_i,e_i]}, & \cdots  \\
    \mathit{[s_1,e_1]},  \cdots, \mathit{[s_i,e_i]}, & \cdots  \\
    \mathit{[s_1,e_1]},  \cdots, \mathit{[s_i,e_i]}, & \cdots
    \end{array}\right],
\end{equation}
where $[s_i,e_i]$ is the start and end time of the $i$-th execution period of a subtask. 
The subtask has an uncertain number of execution periods.
An instance of $ST$ is shown in Fig.\ref{subt}.
The cruise task requires the mobile node to patrol back and forth on the specified path, detect in the designated area, and transmit data to the buoy node.

We extract the task demand based on the $rt$ matrix of each subtask and historical task execution data.
The demand extraction method is not specifically quantified, but there are a variety of ways to realize it, such as LSTM, Particle Swarm Optimization (PSO) algorithm and so on.
$RT$ is input to the demand extraction method to obtain the task demand matrix $TD$, which contains the task demand $td$ of each subtask.
The task demand $td$ can be categorized into three types, corresponding to the three types of tasks, represented as
\begin{equation}
  \begin{aligned}
    \label{td}
    & td_{com} = [ throughput, delay, loss,\dots], \\
        & td_{det} = [ accuracy, coverage, update\_rate, \dots ] ,\\
    & td_{nav} = [ speed, endurance, secrecy, \dots],
  \end{aligned}
\end{equation}
where the three types of subtasks have performance demands in diverse metrics.

\subsection{Slice Schedule}
Slice schedule is responsible for task slices resource schedule according to task demand matrix $TD$ and the global network status.

Global DT core aggregates local DTs to obtain the local state and task scheduling of each node, enabling to obtain the current global network status and predict network trends both temporally and spatially through the simulator.
We obtain the network status matrix $NS = [ns_{1}, \cdots,ns_{i},\cdots, ns_{n}]$  from the global DT core,
where $n$ is the number of all nodes, $ns_{i}$ indicates the state of the $i$-th node, along with the tasks currently executed by the $i$-th node, represented as
\begin{equation}
  \label{nsi}
  ns_{i} = [nodeid, location, speed,\dots, tasks],
\end{equation}
where $tasks$ refers to the set of tasks executed by the node.

When a new task demand matrix $TD$ is produced, TNSD creates task slices corresponding to its subtasks and then conducts resource schedule for the slices based on $NS$. 
TNSD utilizes the specific optimization algorithm to seek the slice resource schedule solution and finally generates the slice resource schedule matrix $RS$ for each subtask as

\begin{equation}
  \label{ra}
  RS = [node_{1},\dots,node_{i},\dots,node_{n}],
\end{equation}
\begin{equation}
  \begin{aligned}
    \label{nr}
    & node_{i} = [location, speed, \dots, role, duration],\\
    & role \in \{source, sink, relay, leader, member,\dots \},
  \end{aligned}
\end{equation}
where $node_{i}$ indicates the state, role, and duration of the $i$-th node in the slice.

We obtain each individual scheduling scheme $ins$ of all nodes by combining the $RS$ of each slice. The individual scheduling scheme $ins_i$ of the $i$-th node is represented as
\begin{equation}
  \begin{split}
    \label{ins}
    ins_{i} &= [task_1,\dots,task_{ni}]\\
    &=\left[\begin{array}{ccc}
      role_{1} & \cdots & role_{ni} \\
      duration_{1} & \cdots & duration_{ni} \\
      \vdots & \ddots & \vdots \\
      status_{1} & \cdots & status_{ni}
      \end{array}\right]\\
                                          \end{split},
\end{equation}
where $ni$ is the number of tasks executed by node $i$. 
The $ins$ indicates the node's task execution sequence, including the role, duration and status in the task, which is utilized to concretely instruct the node to perform the tasks.

We need to determine the network configuration for task slices, such as MAC, routing, power, frequency, etc.
The network configuration is initially assigned based on previous experience and are iteratively evaluated and optimized by the simulator in conjunction with all $ins$.
Finally, the optimal task slice scheme is distributed to each local DT.

\begin{figure}[t]
  \centerline{\includegraphics[width=0.5\textwidth]{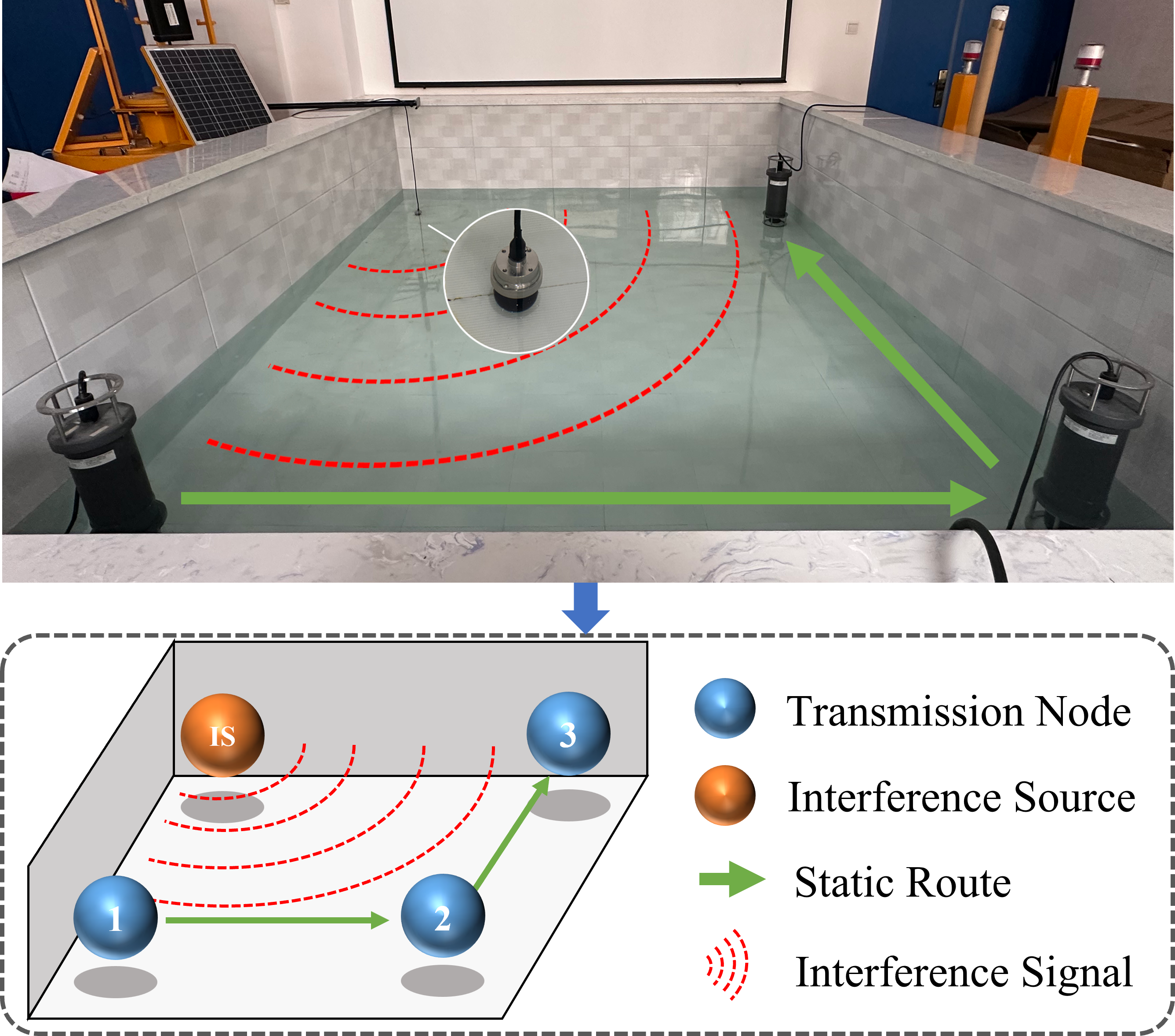}}
    \caption{DT fidelity test scenario. Node $1$ and $2$ are data sending nodes. Node $3$ is the sink node. Interference source sends the interference signal to simulate channel quality degradation.}
  \label{dtctp}
\end{figure}

\begin{table}[t]
  \centering
  \caption{Initial Transmission Parameters}
  \renewcommand{\arraystretch}{1.3}   
  \begin{tabular}{cccc}   
    \hline
    \textbf{Parameters} & \textbf{Values}  &\textbf{Parameters} & \textbf{Values} \\   \hline
    Slot Length & 6 s & Preamble Delay & 0.5 s \\     
    Packet Size & 400 Bytes & Guard Time & 0.05 s \\   
    Sending Rate & 0.03 & Slot Cycle & 3 \\   
    Transmission Rate & 1500 bps & Static Route & 1$\rightarrow$2$\rightarrow$3 \\   
    \hline
  \end{tabular}
  \label{tp}
\end{table}
\begin{table}[t]
  \centering
  \caption{The details of dynamic changes}
  \renewcommand{\arraystretch}{1.3}   \begin{tabular}{ccc}   \hline
  \textbf{Timing} & \textbf{Parameter}& \textbf{Change} \\   \hline
  500s & Sending Rate& from 0.03 to 0.08 \\ 
  1000s & Packet Size& from 400 B to 200 B\\       
  1500s & Slot Length& from 6 s to 4 s\\   
  2000s & Channel Quality& Active Interference Source  \\   
  \hline
  \end{tabular}
  \label{dc}
\end{table}
\begin{figure}[t]
  \captionsetup[subfigure]{font=small} 
  \captionsetup[subfloat]{captionskip=1pt}   
  \subfloat[]{
      \includegraphics[scale=0.78]{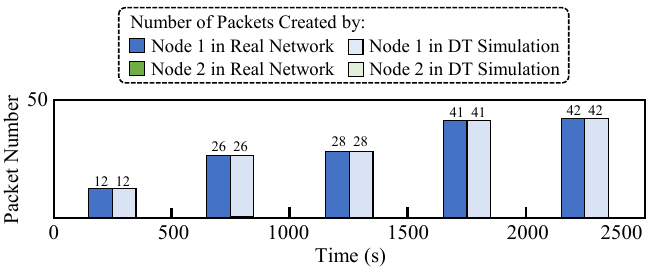}}\\
      [-0.1em]
  \subfloat[]{
      \includegraphics[scale=0.78]{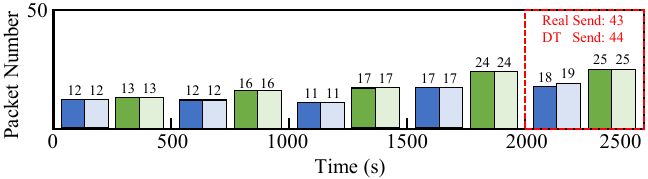}}\\
      [-0.1em]
  \subfloat[]{
      \includegraphics[scale=0.78]{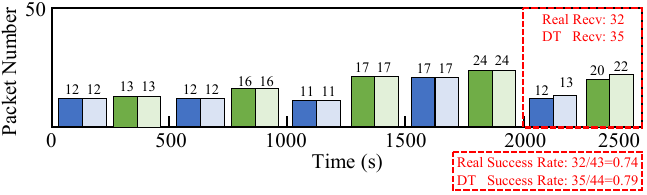}}
    \caption{Number of packets sent/received in each real node and local DT at various time ranges: (a) number of packets sent by Node $1$; (b) number of packets sent by Node $2$; and (c) number of packets received by Node $3$.
      }
  \label{localdt}
\end{figure}
\section{Performance Evaluation}
In this section, we evaluate the performance of our proposed DTNA. 
To our knowledge, there is no comparable research works on underwater digital twin network architecture.
Thus we evaluate the enhancement effect of DTNA on the existing underwater methods.
We evaluate some existing works on underwater resource allocation, MARL algorithms, and multi-task scheduling.
Then we further integrate these works with DTNA and compare them to the original methods to evaluate the performance improvements enabled by DTNA on existing works.
Firstly,
we test the DT fidelity using real devices.
Based on this experimental setup,
we evaluate the timeliness and robustness improvement of RAPD on resource allocation algorithms.
Moreover, we utilize a simulation environment to evaluate the acceleration of CMFD on MARL training, 
and the cost-effectiveness of TNSD on network status obtainment.

For the DT core, we implement the control unit and database using Python and implement the simulator using the improved AquaSim-Tg \cite{AQUA}.
We develop interfaces for the nodes' device, protocol stack, and intelligent methods, connecting them to the control unit to facilitate real-time data acquisition.

\subsection{DT Fidelity Test}

We conduct a lab pool test to evaluate DT fidelity.
As shown in Fig.\ref{dtctp}, we deploy three acoustic modems as transmission nodes and one acoustic modem as an interference source in the pool.
The transmission nodes adopt static routing and Pipelining MAC (PMAC) protocol for communication \cite{pmac}.
The initial main parameters of the nodes are shown in Table~\ref{tp}.
The slot length is determined by the sum of transmission and propagation delay.
We calculate the transmission delay of $200B$ and $400B$ packets to be $1.817s$ and $3.133s$ based on the transmission parameters setup. 
The propagation delay is calculated to be $2s$ using a typical underwater communication distance of $3km$ and a sound speed of $1500m/s$ \cite{3km}.
Finally, we add the transmission and propagation delay, and round up to obtain the corresponding time slot lengths of $4s$ and $6s$ for $200B$ and $400B$ packet lengths.

We conduct dynamic changes in time slot length, packet size, sending rate, and channel quality, in which the channel quality change is simulated by the additional acoustic modem to send interference signal.
The details of dynamic changes are shown in Table~\ref{dc}.
We deploy local DT on each transmission node and global DT on node 3.
We test the DT perception of dynamic changes and verify the DT fidelity by comparing the DT performance simulation results with the real performance.

Fig.\ref{localdt} shows the number of packets sent or received in each real node and local DT, which can infer the data traffic and congestion.
It can be observed that each local DT has a high degree of fitting with the real node in the number of packets sent or received.
Specifically,
the increase in the sending rate at $500s$ maximizes the number of packets sent and causes packet congestion, since there are always packets waiting to be sent at each time slot.
Moreover, the packet congestion in node $2$ also causes it to send more packets created by node $2$ rather than relaying packets created by node $1$.
The packet size change at $1000s$ does not affect the sending strategy or the number to be sent.
However, we can observe that difference in the number of the two types of packets sent by node $2$ increases, indicating that the packet congestion becomes increasingly severe over time.
The slot length reduction at $1500s$ relieves packet congestion and increases packet traffic.

After the channel quality decreases at $2000s$, the DT packet numbers are slightly higher than the real ones. 
This is because the local DT needs to analyze several failed communications to evaluate the packet loss probability.
In addition, we calculate the transmission success rate from the numbers of packets sent by node $2$ and received by node $3$ as shown in Fig.\ref{localdt}(b)(c).
It can be observed that the DT transmission success rate is close to the real value.
In conclusion, the local DT rapidly perceives the dynamic changes and accurately simulates the nodes' transmission packet numbers after each change, achieving a high-fidelity replica of the nodes' transmission.


\begin{figure}[t]
  \centerline{\includegraphics[width=0.48\textwidth]{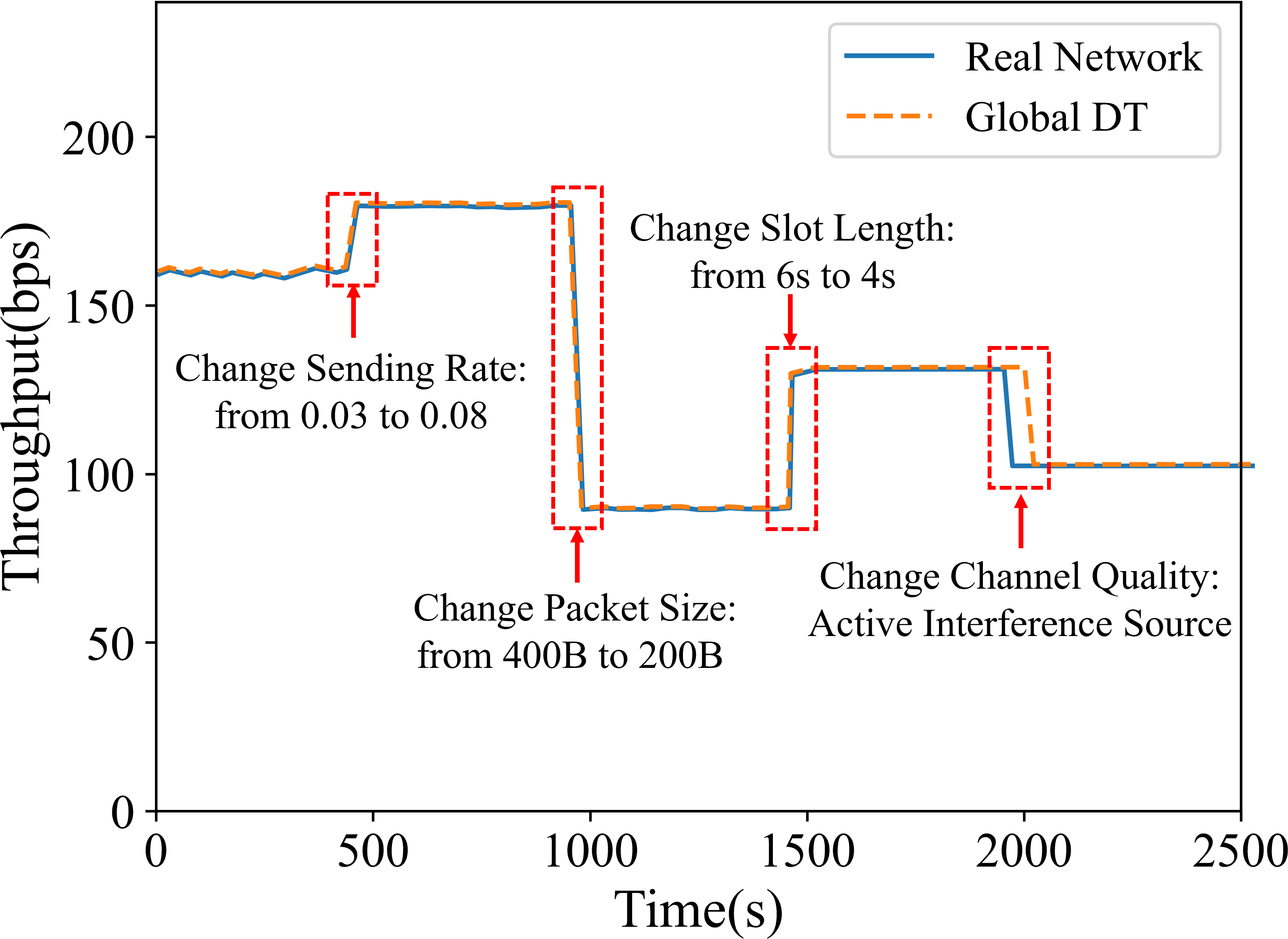}}
      \caption{Throughput of the real network and gloabl DT.}
  \label{globaldt}
\end{figure}

Fig.\ref{globaldt} shows the throughput of the real network and global DT.
It can be observed that the global DT realizes real-time perception of throughput variations to dynamic changes in sending rate, data packet size, time slot length, and channel quality. 
It should be noted that after significantly increasing the sending rate, the throughput increases slightly.
This is because the sending rate of $0.03$ makes the network throughput close to the theoretical upper limit at the current slot length, so increasing the sending rate fails to greatly increase throughput and instead congests the packet queue.
After the channel quality decreases, the global DT's throughput is slightly delayed compared to the real network. 
Similar to local DT, this is because the global DT needs time to evaluate the packet loss probability.
In brief, the global DT realizes the network throughput perception and simulator to each dynamic change, achieving a high-fidelity network replica.

\subsection{Timeliness and Robustness Test for RAPD}

Based on the DT fidelity lab pool test, we conduct test to evaluate the timeliness and robustness improvement of RAPD on the selected resource allocation algorithm.
In the test scenario as shown in Fig.\ref{dtctp}, we shut down node $2$ and allow the node $1$ and $3$ to alternately send $200B$ packets at a $4s$ time slot. 
Moreover, the size of a header-only RTS packet is $30B$ and the highest transmission power of the acoustic modems is $30W$.
The interference source continuously sends random signals at varying power to simulate the interfering acoustic entities.

\begin{table}[t]
  \centering
  \begin{threeparttable}
    \caption{Performance of Three Methods. }
    \renewcommand{\arraystretch}{1.3}     \setlength{\tabcolsep}{1.5pt}     \begin{tabular}{cccccccc}     \hline
        &\textbf{I-Power(W)} & 0.5 & 1 & 2 & 4 & 6 & 8 \\     \hline
        \multirow{2}{*}{\textbf{PC-MAC}} & \textbf{T-Power(W)} & 6 & 6 & 6 & 6 & 6 & 6 \\     & \textbf{BER} & 0.002 & 0.002 & 0.366 & 0.61 & 0.62 & 0.73 \\     \hline
        \multirow{4}{*}{\textbf{PC-MAC-R}} & \textbf{T-Power(W)} & 6 & 6 & 12 & 12 & 18 & 18 \\     & \textbf{BER} & 0.002 & 0 & 0.002 & 0 & 0 & 0 \\     & \textbf{Adjustment Time(s)} & / & / & 8 & / & 8 & / \\         & \textbf{Energy Consumption(J)} & / & / & 43.1 & / & 54 & / \\     \hline
        \multirow{4}{*}{\textbf{PC-MAC-DT}} & \textbf{T-Power(W)} & 6 & 6 & 8 & 12 & 16 & 18 \\     & \textbf{BER} & 0 & 0.002 & 0.004 & 0 & 0 & 0 \\     & \textbf{Adjustment Time(s)} & / & / & 0.1 & 0.1 & 0.1 & 0.1 \\         & \textbf{Energy Consumption(J)} & / & / & 14.5 & 21.8 & 29 & 32.7 \\     \hline
    \end{tabular}
    \label{rapdp}
    \begin{tablenotes}[para,flushleft]
      Note: I-Power refers to the interference source power. 
      T-Power refers to the the transmission node power.
      Adjustment time refers to the time to adjust power to restore communication. Energy consumption refers to the energy to adjust the transmit power plus one successful communication.
    \end{tablenotes}
  \end{threeparttable}
\end{table}

We select the Power Control based handshake-competition MAC (PC-MAC) as the baseline and use PC-MAC-Relink (PC-MAC-R) and PC-MAC assisted by RAPD (PC-MAC-DT) for comparison \cite{pcmac}.
\begin{enumerate}
    \item{\textbf{PC-MAC} allows nodes to send RTS at the highest transmission power at the handshake stage, and the receiving nodes calculate the minimum transmission power of the current communication channel based on the signal strength of the received RTS, which is used as the fixed transmission power for the subsequent communication.} 
    \item{\textbf{PC-MAC-R} also uses the handshake strategy to determine the transmission power. However, PC-MAC-R re-handshakes to determine an appropriate transmission power after the BER exceeds the set threshold, thereby restoring the communication link.}
  \item{\textbf{PC-MAC-DT} leverages local DT to perceive the data and interference signal strength. When the BER is excessive to meet the demand, PC-MAC-DT evaluates the channel quality based on the historical signal strength, and determines the appropriate power by iterative optimization.}
  
\end{enumerate}

We evaluate the adjustment time and energy consumption required by each protocol to conduct resource allocation to restore communication.

Table~\ref{rapdp} shows the test results.
It can be observed that the BER of PC-MAC increases with I-Power.
This is because PC-MAC determines the fixed transmission power in advance during the handshake stage, but the fixed power only adapts to small channel quality fluctuations.
Moreover, the BER of PC-MAC-R is consistently minimized against varying I-Power.
PC-MAC-R can re-handshake to determine an appropriate transmission power to restore communication.
However, PC-MAC-R has a long adjustment time and high energy consumption, this is because the re-handshake requires additional control packets, which increases communication overhead.
In addition,
we can observe that PC-MAC-DT has a short adjustment time and low energy consumption while maintaining a low BER.
PC-MAC-DT can perceive the signal-to-noise ratio of the received packets to evaluate the noise level of the interference source, and utilize the local DT for iterative optimization to select the appropriate transmission power to guarantee communication quality.
The resource allocation of PC-MAC-DT is conducted locally, so there is less time and energy consumption for exchanging additional control information.
PC-MAC-DT selects lower power while guaranteeing communication compared to PC-MAC-R, achieving higher allocation accuracy and saving energy for subsequent data transmission.
In conclusion, PC-MAC-DT outperforms PC-MAC and PC-MAC-R in adjustment time, energy consumption, and allocation accuracy, exhibiting improved timeliness and robustness.

\begin{table*}[t]
  \centering
  \caption{Collection Rate and Training Time of DCMD and DCMD-DT in Four Scenarios}
  \renewcommand{\arraystretch}{1.3} 
  \begin{tabular}{ccccccccc} 
  \hline
  \multirow{2}{*}{\textbf{Scenario}} & \textbf{Number} & \textbf{Number} & \multirow{2}{*}{\textbf{Algorithm}} & \multirow{2}{*}{\textbf{Real Episodes}} & \multirow{2}{*}{\textbf{DT Episodes}} & \multirow{2}{*}{\textbf{Collection Rate}} & \multirow{2}{*}{\textbf{Training Time}} \\
  & \textbf{of AUVs} & \textbf{of SNs} & & & & & \\ 
  \hline
  \multirow{2}{*}{\uppercase\expandafter{\romannumeral1}  }&\multirow{2}{*}{2}&\multirow{2}{*}{20}&DCMD & 40000 &  / & 63.15\%  & $1.1\times10^{5}$h \\ 
  &&&DCMD-DT & 15 & 70000&  62.89\% & $3.78$h\\ 
  \hline
  \multirow{2}{*}{\uppercase\expandafter{\romannumeral2}}&\multirow{2}{*}{3}&\multirow{2}{*}{20}&DCMD & 60000 &  / & 70.68\% & $1.5\times10^{5}$h\\ 
  &&&DCMD-DT & 12 & 90000 &  71.76\% & $3.76$h\\ 
  \hline
  \multirow{2}{*}{\uppercase\expandafter{\romannumeral3}}&\multirow{2}{*}{3}&\multirow{2}{*}{25}&DCMD & 60000 &  / & 71.32\% &$1.54\times10^{5}$h\\ 
  &&&DCMD-DT & 12 & 90000&  71.81\% & $3.76$h \\ 
  \hline
  \multirow{2}{*}{\uppercase\expandafter{\romannumeral4}  }&\multirow{2}{*}{4}&\multirow{2}{*}{25}&DCMD & 80000 & /&  73.24\% &$1.8\times10^{5}$h \\ 
  &&&DCMD-DT & 9 & 120000 &  73.76\%& $3.11$h \\ 
  \hline
  \end{tabular}
  \label{performance}
\end{table*}

\subsection{Acceleration Test for CMFD}
Due to the limitations of the field and equipment, we conduct the simulation test to evaluate the acceleration of CMFD on MARL training by utilizing a simulated environment with actual sound speed data instead of a lab pool.
The simulation environment is implemented by integrating Python and AquaSim-Tg, which can emulate the underwater acoustic channel by bellhop to calculate the propagation loss based on sound speed data.
The computer configurations are given as follows: \textit{AMD 5995WX} CPU, \textit{NVIDIA RTX 4090} GPU.

Similar to the real network in Fig.\ref{fig1}, 
we consider a multi-AUV collaborative data collection task, where multiple AUVs are deployed to collect sensor data from each sensor node (SN) on the seabed.
AUVs require intelligent trajectory planning algorithms to maximize data collection rates and energy efficiency.
Time is equally divided into fixed slots. 
In each slot, each AUV broadcasts state for exchanging collaborative information, and selects the velocity and target SN according to the trajectory planning algorithm.
We consider four data collection scenarios in a 5km$\times$5km area with different numbers of AUVs and SNs as shown in Table~\ref{performance}, where SNs are randomly deployed.

We adopt Data Collection scheme for Multi-modal underwater sensor networks based on Deep reinforcement learning (DCMD) as the baseline and utilize DCMD assisted by CMFD (DCMD-DT) for comparison \cite{EDC}.

\begin{enumerate}

  \item{\textbf{DCMD} is a MADDPG-based multiple AUVs trajectory planning algorithm for underwater data collection. DCMD guides AUVs to select appropriate velocity and target SN considering underwater obstacles and currents, to maximize collection rate and energy efficiency.}
      
  \item{\textbf{DCMD-DT} has the same model design as DCMD. 
  However, DCMD-DT requires each AUV not only to perform data collection but also to observe obstacle locations and environmental data such as ocean currents, sound speed, and noise. 
  Global DT constructs the high-fidelity replica of the data collection scenario based on these observations, providing sufficient DT data and patching the impaired real data.}
              
\end{enumerate}



DCMD interacts with the environment to simulate real execution until the model training converges.
DCMD-DT training alternates between real execution epoch and DT execution epoch.
The episode number of a real execution epoch is determined by the number of AUVs. 
The scenario with more AUVs requires fewer episodes to collect data to construct the global DT.
According to \cite{EDC}, we set the hyperparameters of the DCMD and DCMD-DT as follows: actor/critic hidden layer neuron number $64$, batch size $1024$, discount factor $0.95$, actor/critic learning rate $0.001$.

Table~\ref{performance} shows the episode number of real execution and DT execution required for DCMD and DCMD-DT model training, as well as the collection rate and training time.
It can be observed that the DCMD training time is extremely high, this is because DCMD utilizes real execution for model training and requires a large number of episodes to converge, which indicates the infeasibility of training directly in the real underwater environment.
We can also observe that the training time of DCMD-DT is significantly reduced, because sufficient high-fidelity DT data from global DT accelerates the convergence of the models.
DCMD-DT can construct the global DT with merely a few real execution episodes to collect environmental data, such as current and channel quality.
This enables the utilization of DT execution to assist model training, significantly accelerating the model training.
Moreover, DCMD and DCMD-DT have almost the same collection rate, which verifies the fidelity of the DT data. 

\begin{figure}[t]
  \centerline{\includegraphics[width=0.5\textwidth]{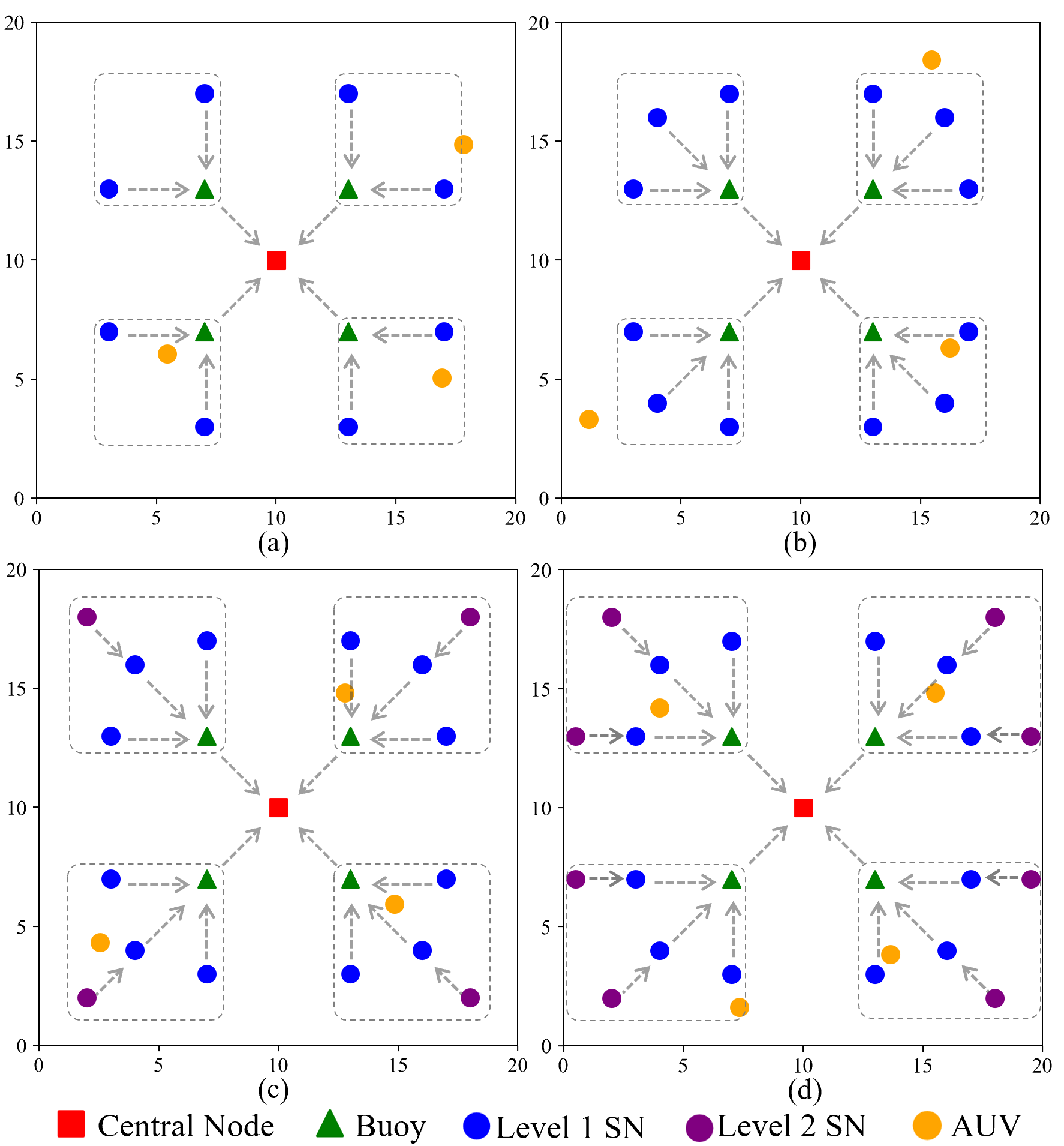}}
        \caption{Multi-task large-scale network scenarios. In the scenarios, the network is divided into four clusters where Buoys are the cluster head and SNs are the member. Central node is the data sink and conducts task scheduling. AUVs are the mobile node and do not belong to any clusters.}
  \label{fc}
\end{figure}
\begin{table}[t]
  \centering
  \caption{Transmission Parameters}
  \renewcommand{\arraystretch}{1.3}   \begin{tabular}{cc}   \hline
  \textbf{Parameters} & \textbf{Values} \\   \hline
  Communication Power & 32 W \\   Electric-Acoustic Conversion Efficiency & 0.5 \\   Schedule Packet Size & 50 Bytes \\   Status Packet Size & 600 Bytes \\   Transmission Mode & BPSK \\   Transmission Rate & 1500 bps \\   Preamble Dealy & 0.5 s \\  Time Slot Length & 7.92 s \\   Slot Cycle & 4 \\   \hline
  \end{tabular}
  \label{para}
\end{table}

In addition, we can observe that the final collection rates of the converged models fail to reach the desired values, because DCMD utilizes a fixed communication strategy to collect data, which ignores the impact of the real ocean environment on communication, resulting in possible collection failures.
CMFD accelerates model training, but cannot fix the inherent flaws of the model itself.

In conclusion, CMFD greatly accelerates model training by generating extensive high-fidelity DT data.
Moreover, CMFD trades off the scarce mobile and communication resources of the distributed nodes with the abundant computational resources of the central nodes, extending the network lifetime.

\subsection{Resource Saving Test for TNSD}
We conduct simulation test to evaluate the time and energy consumption of obtaining the global network status to evaluate the resources saving effect of TNSD.

The main scenario of TNSD is the large-scale, node-heterogeneous networks with multiple tasks.
Therefore, we consider a large-scale network with communication, detection, and navigation capabilities, including multiple types of surface and underwater nodes.
\begin{itemize}
    \item{\textbf{Central Node} serves as the data sink node. It also collects node status and environmental information to issue commands for multi-task scheduling.}
    \item{\textbf{Buoy} serves as a cluster head node. It is responsible for forwarding data to the central node and forwarding commands to the underwater nodes.}
    \item{\textbf{SN} is responsible for monitoring the underwater environment. 
    SNs need to upload data to buoys, and the transmission process may require multi-hop.}
    \item{\textbf{AUV} is responsible for navigation and detection, and can substitute for the data forwarding node when it fails. }
\end{itemize}

We consider four multi-task scenarios in a $20km\times20km$ sea area, containing $1$ central node, $4$ buoys, and varying numbers of SNs and AUVs.
The specific node number and topology of the four scenarios are shown in Fig.\ref{fc}. 
All nodes adopt the same transmission parameters as shown in Table~\ref{para}, where the time slot length is determined by the sum of the maximum propagation delay between any neighboring nodes and the transmission delay of the largest data packet.

\begin{table}[t]
  \centering
  \caption{Consumption of once TGNSO.}
  \renewcommand{\arraystretch}{1.3}   \begin{tabular}{ccccc}   \hline
  \multirow{2}{*}{\textbf{Scenario}} & \textbf{Schedule} & \textbf{Upload} & \textbf{Total Time} & \textbf{Energy} \\   & \textbf{Time} & \textbf{Time}&\textbf{Consumption} & \textbf{Consumption}\\   \hline
  a &  60.96 s & 121.92 s & 182.88 s & 8938.71 J\\   b &  60.96 s & 152.4 s & 213.36 s & 11141.39 J \\   c &  91.44 s & 182.88 s & 274.32 s & 14768.07 J \\   d &  91.44 s & 213.36 s & 304.8 s& 18679.55 J \\   \hline
  \end{tabular}
  \label{fcr}
\end{table}

Multi-task scheduling requires obtaining the global status, including node information such as local state and carrying tasks, network information such as topology and routing, as well as environmental information such as current and channel quality.
Traditional Global Network Status Obtainment (TGNSO) collects and integrates each decentralized node's local information in the following process.
Firstly, the central node broadcasts schedule packets to each buoy, and the buoys forward these packets to the surrounding SNs and AUVs to notify them to upload information.
Then, SNs and AUVs upload the local information to the nearest buoy. 
If they are out of the communication range of the buoy, the SNs and AUVs will upload information via multi-hop transmission through the relays of other SNs.
Finally, the central node collects information from the buoys to obtain the global network status.

We evaluate the time and energy consumption of one TGNSO in four scenarios.
Table~\ref{fcr} shows the test results.
It can be observed that once TGNSO requires considerable time and energy, and the consumption increases with the network scale.
In addition, in order to promptly obtain global network status, it is necessary to interrupt the nodes' task execution for data upload, reducing the network performance.

To evaluate the energy saving effect of TNSD in obtaining the global network status, we conduct further additional setups for the four scenarios.
We set the initial energy of each SN to a random value between 30kJ and 100kJ.
In addition, we import ocean current data to simulate its impact on node drift.
We define that it is necessary to obtain global network status for multi-task scheduling in the following cases:
\begin{itemize}
  {
    \item{Node Drift: when a node drifts out of the communication range of the superior node, the communication is interrupted. The central node needs to obtain global network status to adjust the transmission power and time slot length to restore the connection of the node.}
    \item{Energy Exhaustion: when a node runs out of energy, it fails to send or forward data. The central node needs to obtain the global status to determine which node has run out of energy. For the data relay node, the central node schedules AUVs to replace it for data forwarding to restore communication, while no action is taken for the edge node.}
    \item{AUV Joining: The network topology changes when an AUV joins the cluster network as a data relay node. The central node needs to adjust the node's sending time slot sequence and routing strategy to ensure conflict-free and correct data forwarding.}
          }
\end{itemize}

\begin{figure}[t]
  \captionsetup[subfigure]{font=small} 
  \captionsetup[subfloat]{captionskip=1pt}   \subfloat[]{
      \includegraphics[scale=0.29]{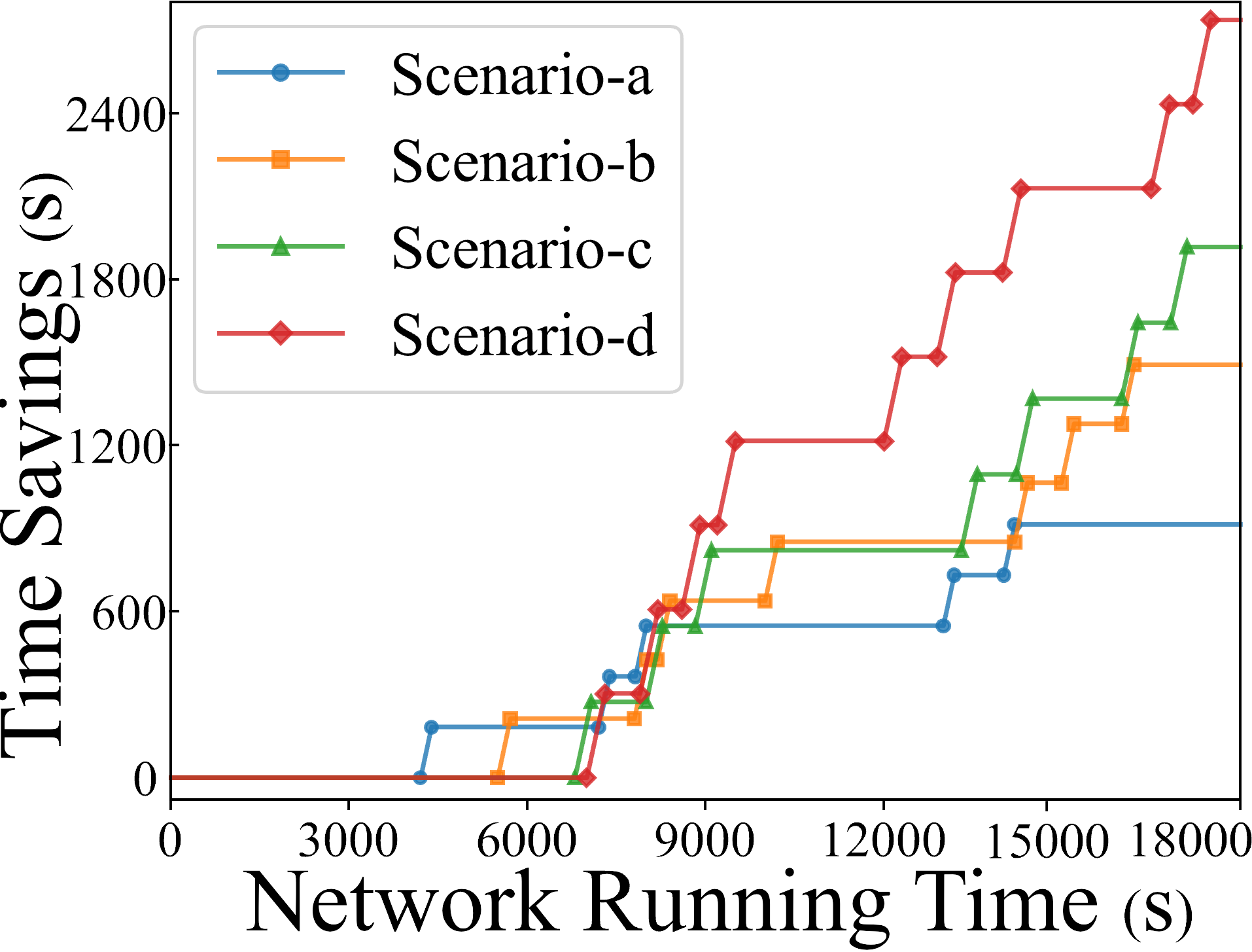}}
  \hspace{-0.2cm}   \subfloat[]{
      \includegraphics[scale=0.29]{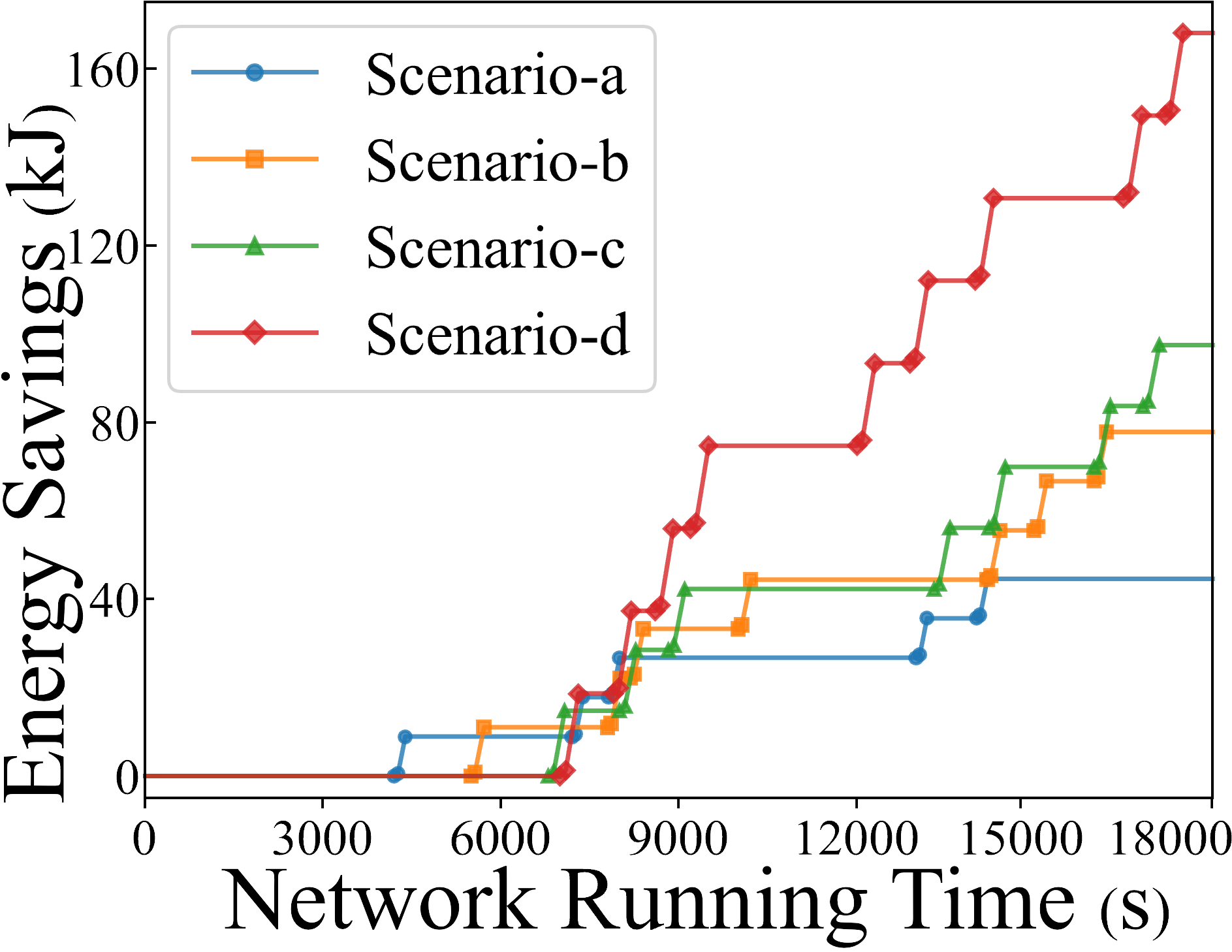}}

    \caption{TNSD compared to TGNSO in terms of resources savings in four different scenarios: (a) time savings; and (b) energy savings.}
  \label{time}
\end{figure}

We deploy TNSD in the central node, and construct the global DT at the initial stage of the network through once TGNSO process.
Then the nodes periodically piggyback their local DT data in normal data packets to maintain the global DT during the network lifetime.

We evaluate the time and energy consumption of TGNSO and TNSD over the network lifetime.
Fig.\ref{time} shows the resource savings of TNSD compared to TGNSO.
It can be observed that as the network running time increases, TNSD saves more and more time and energy.
This is because TNSD can avoid the costly and frequent upload of global network status.
TNSD constructs the high-fidelity network replica in global DT by piggybacking few local DT data, rapidly providing a comprehensive network status in a cost-effective way.
Moreover, TNSD is more effective in saving resources for obtaining the global network status at a larger scale network.
This is because the larger scale network has more nodes to upload local information, increasing the complexity of global network status obtainment, but TNSD can ignore this complexity because of avoiding the frequent TGNSO.

\begin{figure}[t]
  \centerline{\includegraphics[width=0.5\textwidth]{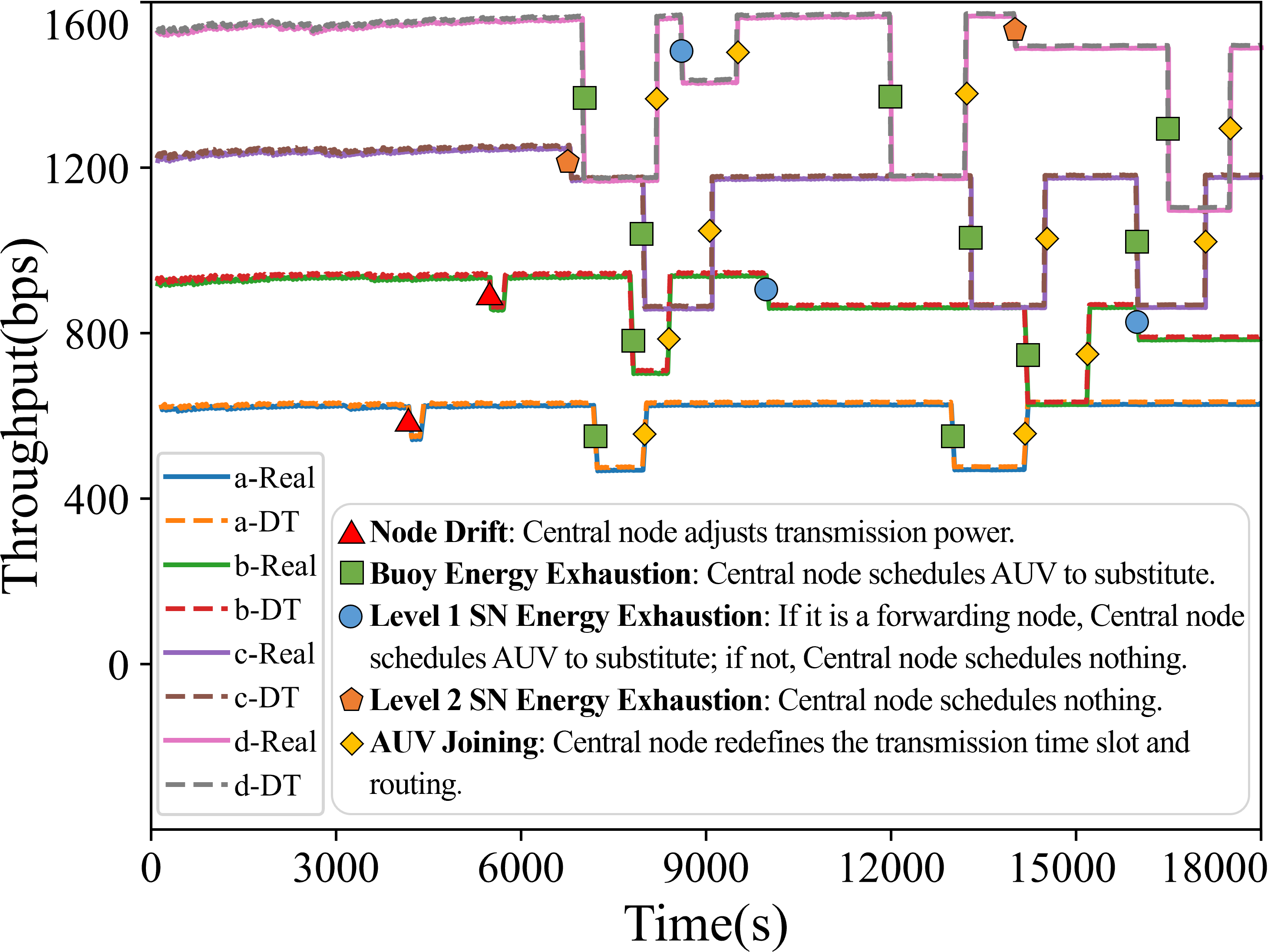}}
  
  \caption{Real and DT throughput comparison. Real network throughput varies by many events, and global DT realizes accurate reproduction through the high-fidelity network replica and trend prediction.}
    \label{th}
\end{figure}

We also compare the real network throughput with the TNSD simulation throughput in the four scenarios.
Fig.\ref{th} shows the results.
It can be observed that TNSD achieves rapid perception of network events, and the simulation throughput has a high degree of fitting with the real network throughput.
This demonstrates that TNSD exhibits timeliness for event perception and accuracy for trend prediction, as well as further verifies the fidelity of global DT. 

In conclusion,
TNSD reduces time and energy consumption to obtain accurate and comprehensive global network status, allowing flexible node scheduling across multi-tasks and extending the network's lifetime.

\section{Conclusion}
In this paper, we present DTNA to enhance the UASNs' environmental adaptability, intelligence, and multifunctionality. 
Specifically, DTNA adopts a layered design to improve the DT replica fidelity and UASN control flexibility.
In local DT, RAPD is designed to rapidly perceive performance variations and iteratively optimize allocation schemes, enhancing real-time environmental adaptability of resource allocation algorithms.
In global DT, CMFD is proposed to accelerate AI model training by patching scarce real data and generating extensive DT data.
In addition, we design TNSD to unify demand extraction for heterogeneous tasks and rapidly provide comprehensive network status at low cost, improving the flexibility of multi-task scheduling algorithms.
Extensive practical and simulation results demonstrate that DTNA accurately replicates real UASNs and effectively improves timeliness and robustness in resource allocation, reduces AI training time, and saves time and energy consumption in multi-task scheduling.

\section*{Acknowledgments}
This work was supported in part by National Key Research and Development Program of China under Grant 2021YFC2803000; in part by the National Natural Science Foundation of China under Grant 62101211, and Grant 62471201.

\bibliographystyle{IEEEtran}
\bibliography{ref}
\vspace{-2em}
\begin{IEEEbiography}
  [{\includegraphics[width=1in,height=1.25in,clip,keepaspectratio]{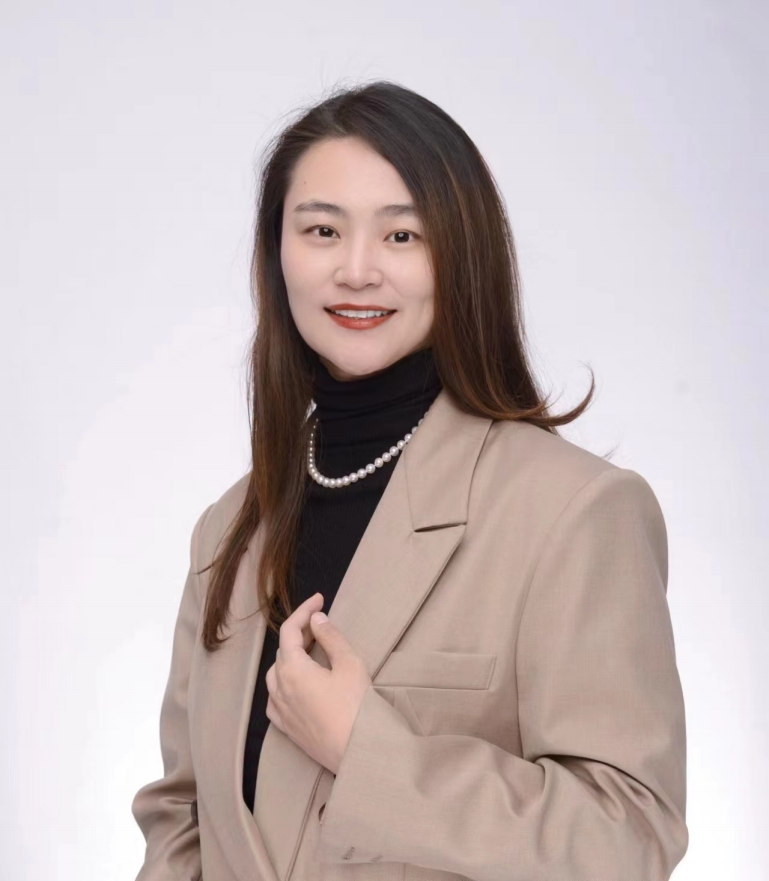}}]{Shanshan Song} (Member, IEEE) received the BS degree (2011) and MS degree (2014) in computer science and technology from Jilin University, China, received PhD degree (2018) in Management science and engineering from Jilin University, China. She was a Post-Doctoral Researcher with the Department of Computer science and technology, Jilin University, Changchun, China. She is currently an associate professor with the Department of Computer science and technology, Jilin University. Her major research focuses on underwater data collection, localization and navigation and machine learning. She serves as the WUWNet' 2023 Publication chair.
\end{IEEEbiography}

\begin{IEEEbiography}
  [{\includegraphics[width=1in,height=1.25in,clip,keepaspectratio]{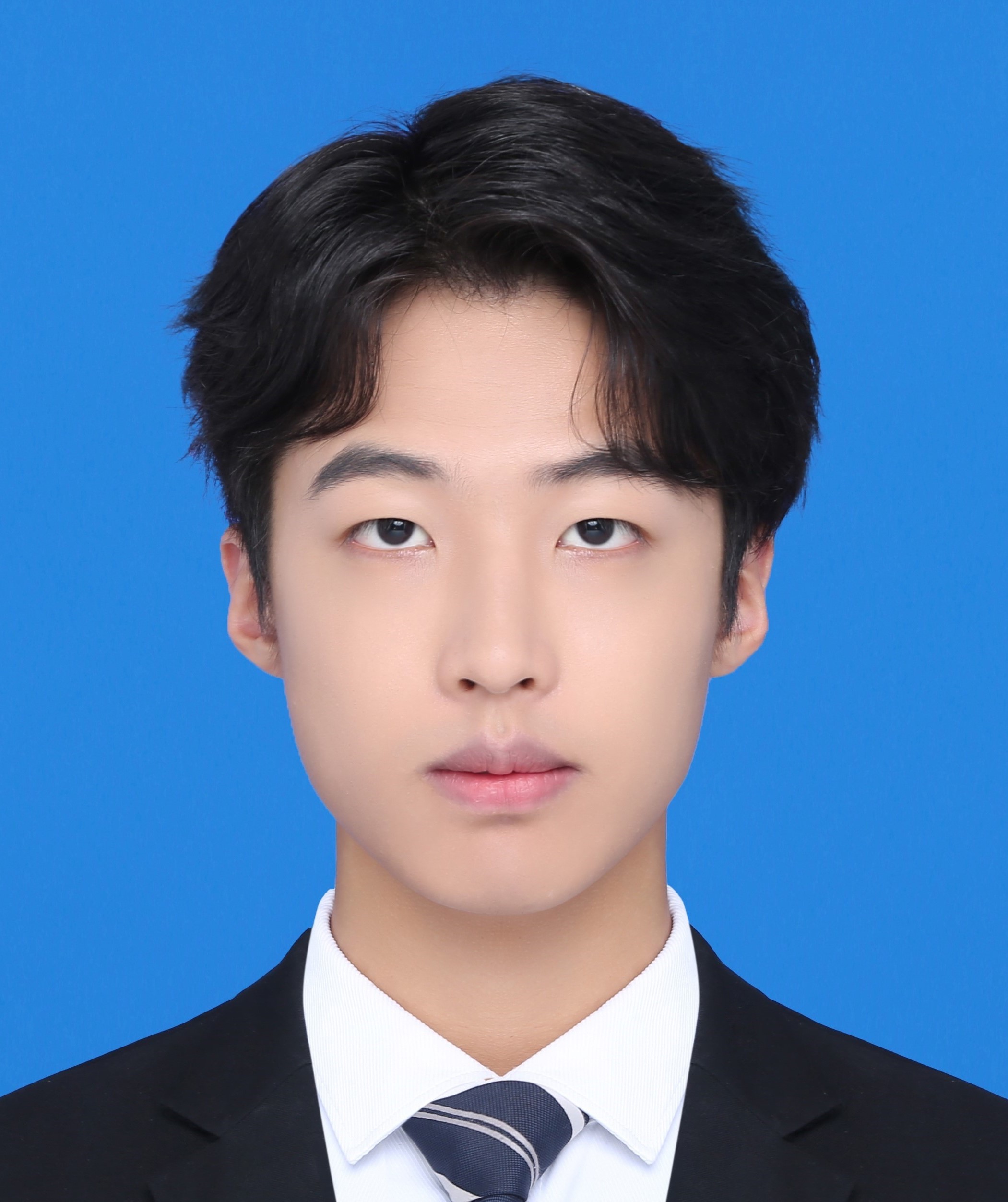}}]{Bingwen Huangfu} received the B.E. degree in computer science and technology from Jilin University, Changchun, China, in 2022. He is currently working toward the PhD degree at the College of Computer science and technology at Jilin University, Changchun, China.
  His current research interests include network architecture, resource allocation, and machine learning for underwater acoustic networks.
\end{IEEEbiography}
\begin{IEEEbiography}
  [{\includegraphics[width=1in,height=1.25in,clip,keepaspectratio]{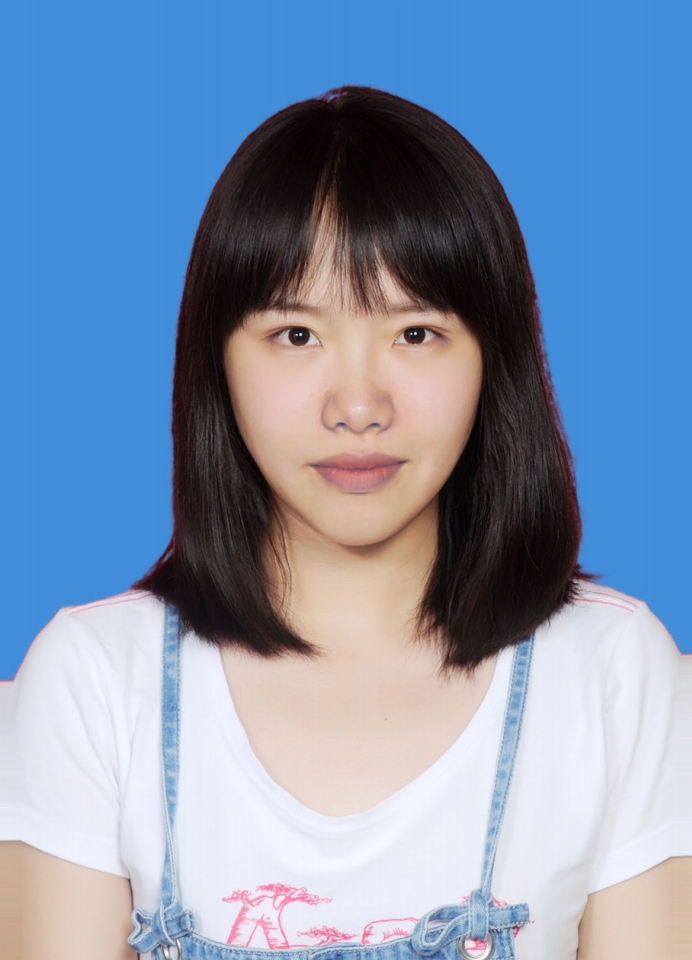}}]{Jiani Guo} received the BS degree (2016) in computer science and technology from Beijing Jiaotong University, Beijing, China, received PhD degree (2024) in Jilin University, Changchun, China. She is currently a Postdoctoral Researcher with the Department of Computer science and technology, Jilin University. Her current research interests include MAC protocols design and performance analysis for underwater acoustic networks.
\end{IEEEbiography}

\begin{IEEEbiography}
  [{\includegraphics[width=1in,height=1.25in,clip,keepaspectratio]{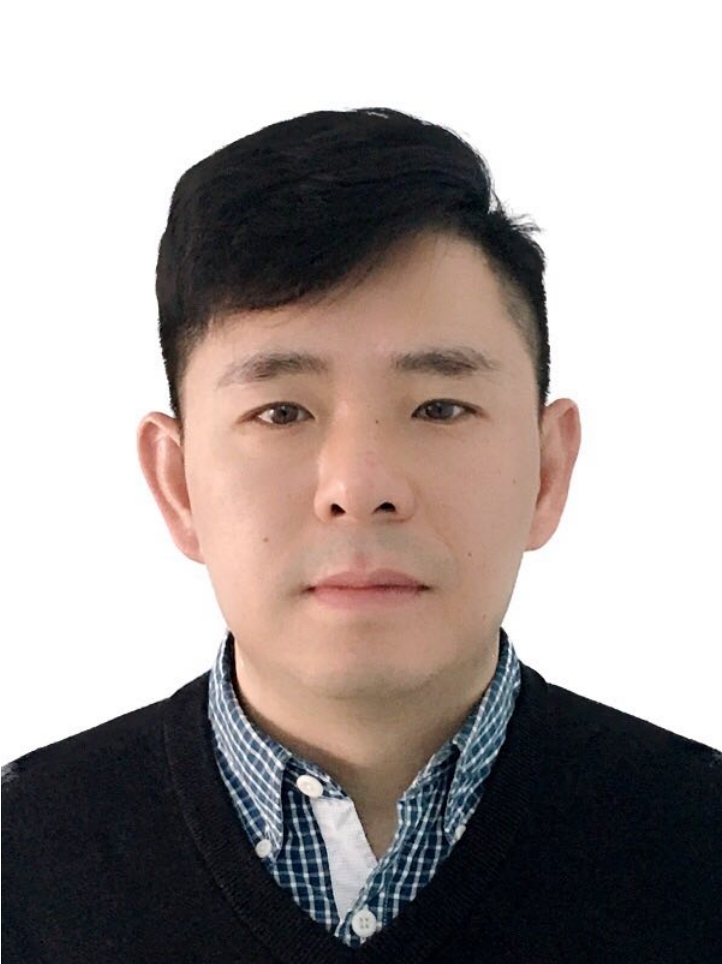}}]{Jun Liu} (Member, IEEE) received the BS degree (2002) in computer science from Wuhan University, China, the PhD degree (2013) in Computer Science and Engineering from University of Connecticut, USA. Currently, he is a professor of the School of Electronic and Information Engineering at Beihang University, Beijing, China, also a part-time professor of the Robotics Research Center, Peng Cheng Laboratory, Shenzhen, China. His major research focuses on underwater acoustic networking, time synchronization, localization, network deployment, and also interested in operating system, cross layer design. He is a member of the IEEE Computer Society.
\end{IEEEbiography}

\begin{IEEEbiography}
  [{\includegraphics[width=1in,height=1.25in,clip,keepaspectratio]{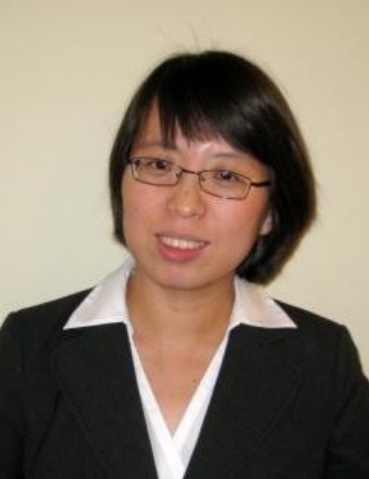}}]{Junhong Cui} received the BS degree (1995) in computer science from Jilin University, China, the MS degree (1998) in computer engineering from the Chinese Academy of Sciences, China, and the PhD degree (2003) in computer science from the University of California, Los Angeles. She was on the faculty of the Computer Science and Engineering Department at the University of Connecticut, Storrs. Currently, she is the professor of the College of Computer Science and Technology at Jilin University, Changchun, China. Her research interests include the design, modeling, and performance evaluation of networks and distributed systems. Recently, her research mainly focuses on exploiting the spatial properties in the modeling of network topology, network mobility, and group membership, scalable and efficient communication support in overlay and peer-to-peer networks, and algorithm and protocol design in underwater sensor networks. She is actively involved in the community as an organizer, a TPC member, and a reviewer for many conferences and journals. She is a guest editor for ACM Mobile Computing and Communications Review and Elsevier Ad Hoc Networks. She cofounded the first ACM International Workshop on UnderWater Networks (WUWNet 2006) and now serves as the WUWNet steering committee chair. She is a member of the IEEE, ACM, ACM SIGCOMM, ACM SIGMOBILE, IEEE Computer Society, and IEEE Communications Society.
\end{IEEEbiography}
\begin{IEEEbiography}
  [{\includegraphics[width=1in,height=1.25in,clip,keepaspectratio]{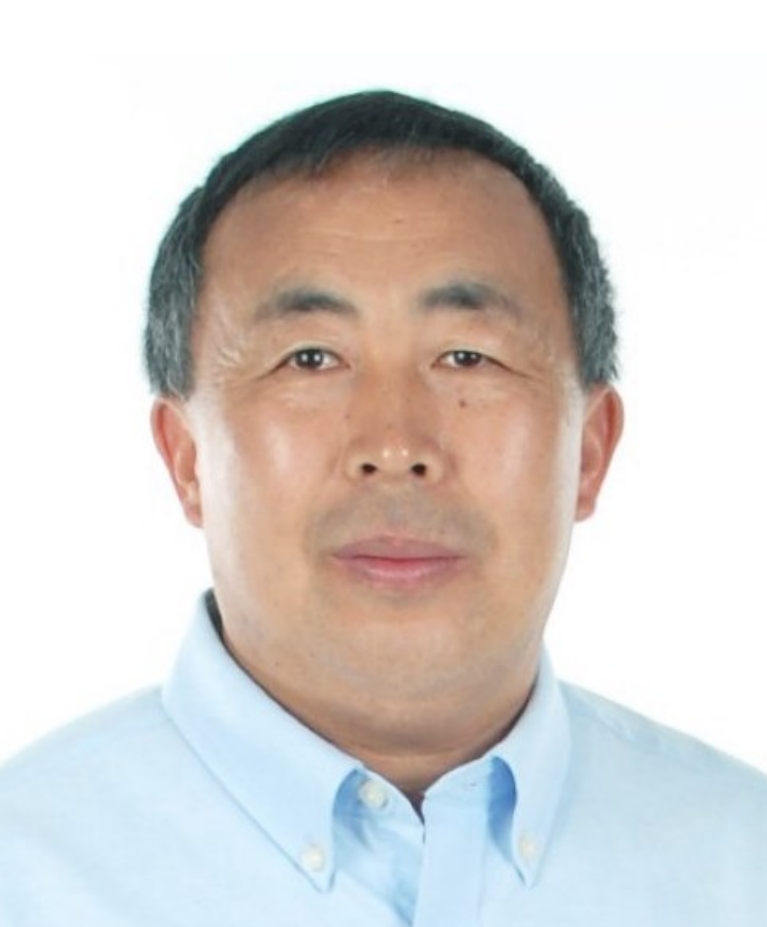}}]{Xuemin (Sherman) Shen}  (Fellow, IEEE) received the Ph.D. degree in electrical engineering from Rutgers University, New Brunswick, NJ, USA, in 1990. He is a University Professor with the Department of Electrical and Computer Engineering,University of Waterloo, Canada. His research focuses on network resource management, wireless network security, Internet of Things, 5G and beyond, and vehicular networks. 
  He received the “West Lake Friendship Award” from Zhejiang Province in 2023, the President's Excellence in Research from the University of Waterloo in 2022, the Canadian Award for Telecommunications Research from the Canadian Society of Information Theory in 2021, the R.A. Fessenden Award in 2019 from IEEE, Canada, the Award of Merit from the Federation of Chinese Canadian Professionals (Ontario) in 2019, the James Evans Avant Garde Award in 2018 from the IEEE Vehicular Technology Society, the Joseph LoCicero Award in 2015 and the Education Award in 2017 from the IEEE Communications Society (ComSoc),
  and the Technical Recognition Award from Wireless Communications Technical Committee in 2019 and an AHSN Technical Committee in 2013. 
  He has also received the Excellent Graduate Supervision Award in 2006 from the University of Waterloo and the Premier's Research Excellence Award (PREA) in 2003 from the Province of Ontario, Canada. 
  He serves/served as the General Chair for the 6G Global Conference'23, and an ACM Mobihoc'15, the Technical Program Committee Chair/Co-Chair for IEEE Globecom'24, 16, and 07, IEEE Infocom'14, IEEE VTC'10 Fall, and the Chair for the IEEE ComSoc Technical Committee on Wireless Communications. 
  He is a registered Professional Engineer of Ontario, Canada, an Engineering Institute of Canada Fellow, a Canadian Academy of Engineering Fellow, a Royal Society of Canada Fellow, a Chinese Academy of Engineering Foreign Member, and a Distinguished Lecturer of the IEEE Vehicular Technology Society and Communications Society. He is the President of the IEEE ComSoc. He was the Vice President for Technical and Educational Activities, the Vice President for Publications, a Member-at-Large on the Board of Governors, the Chair of the Distinguished Lecturer Selection Committee, and a member of IEEE Fellow Selection Committee of the ComSoc. 
  He served as the Editor-in-Chief of the IEEE IoT JOURNAL, IEEE Network, and Peer-to-Peer Networking and Applications.
\end{IEEEbiography}
\end{document}